\documentclass[aps,twocolumn,pre]{revtex4}
\usepackage{graphicx}
\usepackage{latexsym}
\usepackage{amsmath}
\usepackage{amsfonts}
\usepackage{amssymb}
\usepackage{hhline}
\usepackage[]{units}

\usepackage{color}
\usepackage{ulem}

\usepackage[
        plainpages=false, 
        pdfpagelabels, 
        bookmarks=true, 
        bookmarksnumbered = true, 
        bookmarksopen=false, 
        bookmarksopenlevel=3, 
        colorlinks=true,
        citecolor=blue,
        urlcolor=black,
        pagecolor=black,
        linkcolor=black,
        ]{hyperref}

\newcommand{\be}{\begin{equation}}
\newcommand{\ee}{\end{equation}}
\newcommand{\bea}{\begin{eqnarray}}
\newcommand{\eea}{\end{eqnarray}}
\begin{document}
\title{From a melt of rings to chromosome territories: The role of topological constraints in genome folding}





\author{Jonathan D. Halverson$^1$, Jan Smrek$^2$, Kurt Kremer$^3$, Alexander Y. Grosberg$^2$}

\affiliation{$^1$ Center for Functional Nanomaterials, Brookhaven National Laboratory, Upton, NY
11973 USA}

\affiliation{$^2$ Department of Physics and Center for Soft Matter Research, New York University,
4 Washington Place, New York, NY 10003, USA}

\affiliation{$^3$ Max Planck Institute for Polymer Research, Ackermannweg 10, 55128 Mainz,
Germany}

\date{\today}

\begin{abstract} We review \textit{pro} and \textit{contra} of the hypothesis
that generic polymer properties of topological constraints are behind many aspects of chromatin
folding in eukaryotic cells.  For that purpose, we review, first, recent theoretical and
computational findings in polymer physics related to concentrated, topologically-simple (unknotted
and unlinked) chains or a system of chains. Second, we review recent experimental discoveries
related to genome folding. Understanding in these fields is far from complete, but we
show how looking at them in parallel sheds new light on both.
\end{abstract}

\maketitle

\section{Introduction}\label{sec:Intro_short}

Each cell of the human body contains about 2 meters of DNA (46
molecules, 5 centimeters long each on average).  This much DNA is
packed within the cell nucleus with linear dimensions of about five
to ten micrometers.  How is such an extreme folding achieved? In our
view, the natural path to approach the folding of the genome is from
a polymer physics perspective.  The purpose of the present review is
to summarize our recent polymer physics findings
\cite{Vettorel_Grosberg_Kremer_2009,Melt_of_Rings_Statics_2011,Melt_of_Rings_Dynamics_2011,
Linear_contaminants,Comparing_Lattice_and_off-Lattice} and review
their potential implications for the field of genome folding in
light of recent experimental
\cite{van_den_Engh04091992,FISH_Yokota,Chromosome_territories_Cremer,3C,FISH_Gilbert,
4C,FISH_Jhunjhunwala,HiC_Science_2009,Marko_Topo_Constraints_2010,Cremer_FISH_2011,
HiC_for_mouse,HiC_drosophila_Sexton,Cremer_FISH_2012} and
computational achievements
\cite{Rosa_Everaers_PLOS_2008,Everaers_2010_Gene_Colocalization,Arya_PRE,Zimmer_Group_Yeast_Predictive_Model,
Mateos-Langerak10032009,Bohn_Heerman_PLoS_2010,Heermann_chromatin_loops,
Heermann_polymer_loops_chromatin,Heerman_PLOS_2012} (see also review
articles \cite{Chromosome_territories_Cremer_Review_2006,
Mirny_Chromosome_Res_Review,Mirny_CurrentOpinion_2012,Dekker_Chapter_7,vanSteensel_EMBO_review_2011,RosaZimmer_Review,Steensel_Review_2013}).

The very fact that the genome folding problem belongs to the realm
of polymer physics was recognized early on, particularly by B.~Trask
and her co-workers and followers
\cite{van_den_Engh04091992,FISH_Yokota,Hahnfeldt15081993,Sachs28031995,Marko_Siggia01111997,
Ostashevsky01111998,Schiessel_Gelbart_Bruinsma_2001,Ostashevsky01062002,Cook_Marenduzzo_2009}.
More recently this line of research was continued
\cite{Mateos-Langerak10032009,Bohn_Heerman_PLoS_2010,Heermann_Polymer_Loop_Commentary}.
Through these and other works, it is understood that the ``polymer''
in question for genome folding is not naked DNA, but rather the
chromatin fiber -- a complex of DNA with many proteins (histone
complexes) more or less tightly bound to DNA.  Its length is smaller
than 2 meters, but still large enough, on the order of millimeters to
centimeters. To imagine the situation, it is useful to
increase all scales by a factor of $10^{6}$, thus arriving at the
necessity to pack and unpack about one hundred kilometers of a regular
centimeter-thick rope in and out of a delivery truck.

The rope example highlights the role of ``entanglements'', here loosely understood as all
consequences of the fact that two segments of DNA/chromatin cannot cross one another, at least not on
their own. While a hundred kilometers of rope will easily fit in a truck considering
its bare volume, it will be hopelessly tangled if randomly packed.  Any attempt to pull out or
manipulate any particular piece will become almost impossible. Thus, the real problem is not so
much the packing itself, but dealing with the entanglements. Meanwhile, tangling is a very generic
property of any long ``polymer'', completely
\textit{independent of any detail} (e.g.,  spaghetti, rope, wire, fishing line).
Thus the chromatin fiber should be in the same class!
Meanwhile, the cell has to be able to operate on selected parts of DNA with the transcription
factors, RNA polymerase, and all other relevant cell machinery. Roughly speaking, DNA should act
pretty much as a RAM (random access memory) device, allowing easy access to any place.

Although the view of genome folding as a polymer physics problem
seems generally accepted, not much emphasis was placed on the role
of topological constraints and entanglements, with the exception
of \cite{crumpled2,Sikorav_Jannink_1994,Rosa_Everaers_PLOS_2008}.
In Ref. \cite{crumpled2}, the hypothesis of the so-called crumpled globule
(later called also fractal or loopy globule) was formulated as a
possible resolution to the tangling problem of chromatin. The central
idea was the connection between topological simplicity, the lack of
knots, and spatial self-similarity. The estimates of the work
\cite{Sikorav_Jannink_1994} suggested a limited role of topological
enzymes in resolving the conundrum of chromatin tangling.  And the
work \cite{Rosa_Everaers_PLOS_2008}, independently of the much
earlier work \cite{crumpled2}, arrived at the conclusion that topological constraints
play the central role in the whole of the genome
folding problem. In this review, we place topology at the center
stage
\cite{Rosa_Everaers_PLOS_2008,Vettorel_Grosberg_Kremer_2009,PhysicsToday_Image,Rosa_Everaers_Looping2010,
Everaers_2010_Gene_Colocalization,Arsuaga_chromosomes,Stasiak_chromosomes,Heermann_polymer_loops_chromatin,
Bohn_Heerman_Topological_Loop_Repulsion,Heermann_chromatin_loops}.
Of course, we will build on the significant body of knowledge on
polymer topology as described, e.g., in the textbooks
\cite{pgdg,DoiEdwards,RubinsteinColby,RedBook} and even a popular
book \cite{GiantMolecules}.

Interestingly, polymer topology, as we will see, naturally brings together two aspects of genome
folding which were traditionally discussed separately, namely, the polymer physics aspects and
the self-similarity aspects.  The latter has been discussed many times in the chromatin
literature, and there are indications of both a fractal structure in the cell nucleus interior
\cite{Chromatin_Fractality_Early,Chromosome_Fractality_Scattering,Chromosome_Fractal_Review_Bancaud,
Chromosome_fractality_experiment_Bancaud} and of non-classical diffusion of either particles or
the ends of chromatin fiber inside the nucleus \cite{Sedat_1997,Cremer_brothers_dynamics_1999,
Chromosome_fractality_experiment_Bancaud,Dynamics_Golding,Dynamics_Banks, Gasser_2001,Wirtz_2004,
Power-Law_Rheology_2005,Gratton_2005,
Meshorer_2006,Belmont_CurrentBiology_2006,Kanger_NanoLetters_2007,
Discher_PNAS_2007,Dynamics_Bronstein,Wirtz_BioPhysJ_2011,Shiva_2012,
Gratton_2012,Theriot_2012,Garini_2012,Garini_2013,Chromatin_Transitions_Cell_2013}. We here focus
on the attempts to understand these elements of self-similarity based on polymer topology.

The plan of our review is as follows.  We begin Section
\ref{sec:primer} where basic facts and terminology about genome
folding is briefly summarized, mostly for a physicist or a chemist
reader. The central piece of this primer is Table
\ref{tab:examples_of_organisms} which summarizes quantitative information
about genome packing across biological realms. In Section \ref{sec_sub:importance_of_topology}
we formulate the physics view of the subject based on polymer systems
with topological constraints.
We explain what the
topological constraints are and why we believe them to be of
decisive importance. We then formulate in Section
\ref{sec_sub:why_melt_why_rings} the specific workhorse model of our
approach, a melt of ring polymers, estimate the relevant polymer parameters of
chromatin fiber \ref{sec_sub:estimate_entanglement_length}, and
compare chromatin problems to those known in polymer rheology
\ref{sec_sub:rheology}.

The rest of the work presents at least two interweaving streams. The
fact of the matter is that although we argue the melt of rings to be
a relevant model for chromatin folding, the model itself is by no
means completely understood.  We, therefore, combine the review of
efforts to understand a melt of rings and related topological polymer
models with the review of applications that these polymer ideas find for
genome folding. Specifically, Section
\ref{sec:melt_of_rings_and_polymer_models} is mostly about the
model: we review formulation of the model
(\ref{sec_sub:model_and_question_formulation}), the existing (rather
controversial and inconclusive) theoretical approaches
(\ref{sec_sub:theoretical_approaches}), mathematical space-filling
curves (\ref{sec_sub:Peano}), as well as simulation data for the melt
of rings (\ref{sec_sub:simulation_data_rings}) and for other related
models (\ref{sec_sub:simulation_of_other_models}).

We then move on to compare physics insights against experimental
data on chromatin.  We begin in Section
\ref{sec:qualitative_territories} with qualitative observation,
namely, chromosome territories (see the original work
\cite{Chromosome_territories_Cremer}, and the review
\cite{Chromosome_territories_Cremer_Review_2006} with its
many references, and particularly the very readable account
\cite{Territories_Meaburn_Misteli_Nature_News_Views_2007}). The
center of our argument there is Fig. \ref{fig:political_maps}
which demonstrates how topological constraints on polymers yield
a natural simple explanation of territories.  With that in mind, we
move on in Section \ref{sec:territories_quantitative} to
quantitative data. We first review in Section
\ref{sec_sub:Modern_experimental_techniques} modern experimental
methods known under abbreviated names FISH
(\cite{van_den_Engh04091992,FISH_Gerhard_Hummer_1,FISH_Gerhard_Hummer_2,FISH,FISH_Gilbert,FISH_Jhunjhunwala,FISH_Yokota,Mateos-Langerak10032009,Territories_Meaburn_Misteli_Nature_News_Views_2007,FISH_BOOK,Cremer_FISH_2012,Cremer_FISH_2011})
and ``C''
(\cite{3C,4C,HiC_Science_2009,HiC_Yeast,HiC_for_mouse,HiC_drosophila_Sexton};
see also simple summary in \cite{Nature_Methods_NV_Story_of_C}). The
detailed comparison is performed in Section
\ref{sec_sub:contact_probability} for contact probability data and
in Section \ref{sec_sub:subchain_sizes} for subchain sizes.  This
raises new theoretical questions which we address in Section
\ref{sec_sub:theoretical_issues}.  We conclude with a brief discussion
of dynamics in both chromatin and model polymers in Section
\ref{sec:Dynamics} and the possible role of DNA sequence in Section
\ref{sec:Sequence}.

\section{Genome Folding: A Primer}\label{sec:primer}

An overview of the degree of DNA compaction across biological realms
is given in Table \ref{tab:examples_of_organisms}.
By far the densest packing of DNA is achieved in viruses. However,
viruses are very special. First, their genomes are very short. Their
DNA (or RNA) is ``stored'' and nothing happens to it until the virus
infects a cell, i.e., its genome unpacks into the cell.  The study
of viral DNA or RNA, however interesting in its own right (recent
reviews and references can be found in \cite{Review_Viral_DNA} and
\cite{Italian_group_Review_Micheletti_2011}), can teach us
relatively little about packing of genomes in higher organisms.

In prokaryote bacteria cells, biologically the next simplest level,
the DNA  is located in the so-called nucleoid (see, e.g.,
\cite{Nucleoid,Nucleoid_Review}) where it appears to be rather
loosely associated with proteins. Surprisingly little is known about
these proteins and about nucleoid structure in general. However,
see the concise review by Gruber~\cite{MukBEF_Mini_Review_2011} and references
therein. We will not consider bacterial genomes here, although the main
ideas about the role of topology might still be applicable
\cite{Caulobacter_Schiessel_2011}.

Cells of higher organisms are called eukaryotic. They have nuclei
where DNA is packed in tight association with histones and other
proteins. In this work we focus exclusively on eukaryotes.
The hierarchical organization of the genome in eukaryotic cells is
described in many textbooks and on the web, typically accompanied by
beautiful cartoons (e.g.,
\cite{Alberts_Book,Hierarchial_DNA_on_the_Cover,Chromosome_remodeling_web_site}).

The first level of hierarchy is well established:  double stranded
(ds) DNA is wound around histone octamers forming so-called nucleosomes.
This ``beads on a string'' chain of nucleosomes is known as the $10$ (or
$11$) $\mathrm{nm}$ fiber. A lot is known about this level of
organization, including the detailed structure of nucleosomes (see
recent summary in \cite{Nucleosome_structure}) and the stems of
their linkers (see, e.g.,
\cite{Everaers_fluctuations_above_nucleosome_scale_2011}).
A length of dsDNA composed of roughly $147$ base pairs is
wrapped around a histone
octamer in each nucleosome, and segments of roughly $10$ to $100$
(about $40$ in average) base pairs form linkers between these
nucleosomes. Whether the position of nucleosomes along DNA and the
corresponding linker lengths are random, dictated by the maximal entropy
principle of statistical mechanics, or determined by the underlying
DNA sequence and form a special code is an extensively studied and
hotly debated subject (see, e.g., \cite{Nucleosome_Positioning}).

The next level of hierarchy is known as the \unit[30]{nm} fiber. Its
existence is easy to understand: since the linker DNA between
nucleosomes is shorter than the persistence length ($l_p \approx
\unit[150]{ bp}$), the chain of nucleosomes is expected to form a
zig-zag with characteristic width close to \unit[30]{nm}. The fact
of the matter is that such a \unit[30]{nm} fiber is neither
particularly rigid nor very well defined \cite{Schiessel2009}, its
prominence under \textit{in vivo} conditions of the cell is
doubtful.  The idea of its importance seems to be losing popularity
\cite{Hansen_EMBO_2012,Dekker_Chapter_7}.

Just above a few tens of nanometers or above $10^3$ base pairs, there
is presently a big gap in our knowledge and understanding. A good
expression of it is given by
Meyer et al.~\cite{Everaers_fluctuations_above_nucleosome_scale_2011}, who
argue for the existence of a cross-over scale.  Below that
scale the system is pretty rigid, its elements (such as nucleosomes
and linkers) can be crystallized, and their structures are fully
determined. By contrast, well above this scale fluctuations become
important. This is the realm of statistical soft matter physics. In
this range, on the scale of about $10^4$ base pairs and higher, the
overall architecture of the genome in 3D space is not well understood
and its features are only now starting to become known. Our present
review is intended to contribute to this process.

Thus, our subject matter is the folding of chromatin fiber -- an
entity which is not perfectly defined. It definitely has the
\unit[10]{nm} fiber at its core, and it may be somewhat more
organized. Its physical properties will be discussed below.

An indispensable part of the chromatin fiber and its properties is a
multitude of different proteins.  Some of them are histones, and
some of the histones are constituents of nucleosomes.  Other
histones, such as H1, are attached to linker DNA.  Many other
non-histone proteins are also involved, such as cohesins and
condensins, and despite their suggestive names their function apparently
is not restricted to gluing pieces of chromatin fiber together and
maintaining the compactness \cite{Bloom_Kerry_Review_2010,Cell_2013_Proteins}  (see also \cite{Cell_2013_Proteins_NV}).

On the highest level of the nucleus as a whole, chromatin is an
important part of many processes through the cell cycle.  We here
focus on the interphase nucleus, the stage when the cell does not divide
and chromosomes stay swollen inside the nucleus.
For differentiated cells, some part
of DNA, which is not transcriptionally active, is packed somewhat
more tightly in so-called heterochromatin.  The other part, called
euchromatin, is less densely packed and it is involved in transcription
of those genes which have to be expressed in the given cell. The
genes in heterochromatin are silenced through either histone
methylation or interactions with the so-called short silencing RNA
(see, e.g., \cite{Alberts_Book}).  In either case, the placement of
any particular part of DNA into either hetero- or euchromatin is
inheritable via epigenetic mechanisms.  Thus, the above mentioned
analogy with RAM is restricted to euchromatin only.  Nevertheless,
for our consideration it is good enough because a significant
part of DNA has to be easily accessible for bulky processes such as,
e.g., homologous recombination, which would be next to impossible if
the DNA were heavily tangled.

Of course, genome folding and organization is not a static
phenomenon.  Cells live, and the cell nucleus is the place of diverse and
incessant activities. In this review, we will mostly consider
so-called interphase, which is (usually) the longest stage of the
cell cycle. During interphase, chromosomes are (relatively) swollen
or decondensed and they occupy most of the volume inside the nucleus. In that
time, proper genes (i.e., the ones which have to be expressed in the
given cell type) are transcribed into RNA for subsequent protein
synthesis. It is believed that interphase chromosomes are
structurally and spatially organized to help control gene
expression.  At the end of interphase, the cell prepares itself for
division.  This process involves a quite dramatic spatial
rearrangement and leads to the formation of highly condensed mitotic
chromosomes in which transcription appears to be switched off.  Our
consideration concentrates on interphase chromatin.

\begin{table*}
\caption{Examples of the amount of DNA and its packing
characteristics in several different organisms. ``Domain'' means
where the genome is located, which is the virus capsid in the case of
bacteriophages, the cell nucleoid for prokaryotes (bacteria),
and the cell nucleus in the case of eukaryotes (exemplified here by
yeast, drosophila, chicken, mouse and human), and $D$ is the
corresponding diameter.  $N$ is the genome length in base pairs.  $L=bN$
(e.g., for a virus or for a haploid cell, having one copy of DNA) or
$2bN$ (for a diploid cell, having two copies of DNA) is the contour
length of stored DNA, where $b = \unit[0.34]{nm}$ is the length per
base pair. Note that the genome length includes the total length of
DNA, i.e., both genes and the non-coding DNA (exons and introns). For
instance, yeast and human are very nearly the same as far as the
gene length is concerned, but yeast has almost no introns, while we
have many, which is why our genome is two orders of magnitude
larger. The volume fraction of DNA is calculated by assuming that
the DNA double helix is a cylinder of diameter \unit[2]{nm} and
length $L$, which corresponds to $(\pi/4) (\unit[2]{nm})^2
\times (\unit[0.34]{nm /bp} ) \approx \unit[1]{nm ^3/bp}$.
Physically more relevant is the volume fraction of DNA together with
tightly bound proteins.  For the bacterial nucleoid, such a number
is hard to estimate. For eukaryotes, the volume of DNA with histones
can be defined in a number of ways. The most conservative estimate
is based on mass-spectrometry data which suggest that histones
increase the mass (and, given their elementary composition, also the
volume) of DNA by about a factor of $2$. The most aggressive
estimate starts with a typical length of the spacer between
nucleosomes as \unit[80]{bp} or so which is \unit[30]{nm}.  These
spacers are shorter than the persistence length of \unit[150]{bp},
so roughly the system goes as a zig-zag with nucleosomes at the
turns. An upper limit of the excluded volume (virial coefficient) of
two such zig-zags can be estimated as that of a cylinder with
diameter \unit[30]{nm}. There are about \unit[200]{bp} per
nucleosome, which corresponds to the distance along the axis of this
cylinder of about one nucleosome, which is \unit[10]{nm}. This
yields a volume per one base pair of $(\pi/4) (30 \mathrm{nm})^2
\times [10 \ \mathrm{nm} /200 \ \mathrm{bp}] = 30 \
\mathrm{nm}^3/\mathrm{bp}$, which is a factor of $30$ greater than
bare DNA.  The latter most aggressive estimate is also in agreement
with that given in the Supplemental Information of the work
\cite{HiC_Science_2009}.  See a more detailed and nuanced discussion
of all these numbers on the web site \cite{BioNumbers_WebSite} and
in the book \cite{CellBiology_by_Numbers}.
\label{tab:examples_of_organisms}}
\begin{tabular}{||l||c|c|c|c|c|c||}
\hhline{|t:=:t:=t=====:t|}
Organism & Length & Diameter &  & Volume fraction &
\multicolumn{2}{c||}{Volume fraction}  \\
&  of genome, & of domain, & $L/D$ &  of DNA & \multicolumn{2}{c||}{including proteins}  \\
& $N, \ \mathrm{bp}$ & $D, \ \mu \mathrm{m}$ &&& lower & upper \\

\hhline{||-||------||} Bacteriophage (T4) & $1.7 \times 10^5$ & $0.05$ & $\sim 10^{3}$ & $\sim 50 \%$ & \multicolumn{2}{c||}{not applicable}  \\

\hhline{||-||------||}  E. coli & $4.6 \times 10^{6}$ & $1$ & $1500$ & $\sim 1 \%$ & \multicolumn{2}{c||}{not known}  \\

\hhline{||-||------||}  Yeast, haploid & $1.2 \times 10^{7}$ & 2 & $\sim 2 \times 10^{3}$ & $\sim 0.3 \%$ & $\sim 0.6 \%$ & $\sim 10 \%$ \\

\hhline{||-||------||} Drosophila, diploid & $1.5 \times 10^{8}$ & 10 & $\sim 10^{4}$ & $\sim 0.05 \%$ & $\sim 0.1 \% $ & $\sim 1.5 \% $ \\

\hhline{||-||------||}  Chicken, diploid & $1.2 \times 10^{9}$ & 5 & $\sim 2 \times 10^{5}$ & $\sim 4 \%$ & $\sim 8 \%$ & $\sim  100 \%$ \\

\hhline{||-||------||}  Mouse, diploid & $2.8 \times 10^{9}$ & 9 & $\sim 2 \times 10^{5}$ & $\sim 1 \%$ & $\sim 2 \%$ & $\sim 30 \%$ \\

\hhline{||-||------||}  Human, diploid & $3.3 \times 10^{9}$ & 10 & $\sim 2 \times 10^{5}$ & $\sim 1 \%$ & $\sim 2 \%$ & $\sim 30 \%$ \\

\hhline{|b:=:b:=b=====:b|}

\end{tabular}
\end{table*}

\section{Polymer Physics Picture}\label{sec:polymer_physics_picture}

\subsection{Polymer packing and topology}\label{sec_sub:importance_of_topology}

Approaching the polymer physics picture of chromatin folding, we
should first realize that chromatin fiber as a polymer is packed
fairly densely in the nucleus.  The volume fraction of DNA itself in
a typical human nucleus is as high as about a percent (with 3.3
billion base pairs in the genome, two copies of the genome in a
diploid cell, $b=0.34 \ \mathrm{nm}$ of length per base pair, double
helix diameter of \unit[2]{nm}, and typical cell nucleus diameter of
$\unit[10]{\mu m}$, the volume fraction is $1.2\%$ -- see Table
\ref{tab:examples_of_organisms}; see also web site
\cite{BioNumbers_WebSite} and the book
\cite{CellBiology_by_Numbers}). In making such an estimate, one
should keep in mind the tremendous variability of biological
circumstances.  For instance, depending on cell type and conditions,
the diameter of a human cell nucleus can easily vary by a factor of
$2$, from about $6 \ \mu \mathrm{m}$ to $12 \ \mu \mathrm{m}$. This
increases the possible DNA volume fraction by almost an order of
magnitude, making it as high as about $10 \%$. A more meaningful
number, perhaps, would be the volume fraction of the chromatin fibers.
Its exact value depends on the exact definition, two of which are
given in the last two columns of Table
\ref{tab:examples_of_organisms}.  Such estimates also vary
significantly between different types of cells and different
organisms. For instance, the volume fraction of chromatin in simpler
eukaryotes such as yeast is twice or more smaller than in humans. As
usual in biology, there are many different cases, but it appears
that the volume fraction of chromatin is pretty high in all cases.
By the standards of polymer physics, chromatin is not quite as dense
as a melt, but it is a concentrated solution, which should be
thought of as a melt of blobs \cite{pgdg,RubinsteinColby,RedBook}.
Experimentally the properties of concentrated polymer solutions and
those of melts are rather similar without any significant qualitative
difference. For both, the catastrophic tangling of long chains, as in
equilibrated melts/solutions, always leads to macroscopic relaxation
times, which is avoided for chromatin in all cases. In our opinion,
this suggests that genome folding should also be considered from
a polymer physics perspective. Surely, many aspects of chromatin have
very little to do with polymer physics, but one should expect that
the very basic motive -- \textit{being dense while still avoiding
tangling} -- also deserves generic polymer physics considerations.
In our opinion, understanding this is presently one of the main
challenges of the physics of the cell nucleus.

In this context, the most striking observation is that of chromosome
territories \cite{Chromosome_territories_Cremer}.  This observation
is in sharp conflict with the well established observation for
dense and semidilute polymer systems: if chromatin fibers are
considered as regular polymer chains then in a dense system, such as a melt or
concentrated solution, they should interpenetrate, and entangle
if they are in equilibrium. Therefore, already the very fact of
territorial segregation between chromosomes is consistent with the
idea that chromatin fibers somehow avoid tangling. The hypothesis
that DNA in the nucleus might be pretty densely packed while still
avoiding catastrophic tangling was formulated almost 20 years
ago \cite{crumpled2,crumpled_DNA_Russian} based on the preceding
paper \cite{crumpled1}, where the idea of the so-called ``crumpled''
globule was formulated in the context of the dynamics of the coil-globule
transition.  At about the same time, the role of DNA topology and
the corresponding enzymes were also discussed in the work
\cite{Sikorav_Jannink_1994}. More recently but independently from
\cite{crumpled1,crumpled2}, the role of DNA topology was strongly
advocated by Rosa and Everaers \cite{Rosa_Everaers_PLOS_2008}.

When speaking of polymer topology, we mean all consequences of the fact
that two pieces of a real polymer, such as dsDNA or a chromatin
fiber, cannot pass through one another, at least not without special
topo-enzymes.  For polymers without ends, such as rings, the topological
constraints are strict and are topological in the rigorous
mathematical sense of the word.  For open end polymers, topological
constraints are always a matter of time scale.  Rosa and
Everaers~\cite{Rosa_Everaers_PLOS_2008}
provide rather convincing estimates suggesting that on the time
scale of a cell life for higher organisms, topological constraints of
a chromatin fiber should be regarded as permanent.  At the same
time, there are also temporary loops in chromatin fiber every time
that two loci come close to one another in space.  These contacts
and, therefore, loops are detected \textit{en masse} by ``C''
experiments (see below Section
\ref{sec_sub:Modern_experimental_techniques}).  It is important to
realize that these loops do not form topological interactions with one
another as long as their being a loop (i.e., contact between the
ends) is subject to relaxation on the time scale of interest.

\textbf{\textit{Important note on terminology:}} The word ``topology''
seems to be getting over-used in the field of genome folding (as in
many other fields). For instance, people talk about ``topological
domains'' in chromosomes \cite{Nature_2012_C_methods_Many}.  Our use
of ``topology'' follows the polymer physics tradition and, as we
said, means all consequences of the fact that a polymer cannot cross
itself. For our workhorse model of polymer rings, topology will
have a strict mathematical meaning.

Recent experimental work \cite{HiC_Science_2009} appears to be consistent with
the theoretical picture that in the range between $0.7$ and $7$ million base pairs the chromatin fiber is organized as a crumpled globule. In fact,
the authors of the experimental work \cite{HiC_Science_2009} prefer to call this state a fractal instead of crumpled globule, which is a pure terminological discrepancy (i.e., fractal and crumpled globule mean exactly the same thing).  The experimental work strongly motivated closer theoretical scrutiny.  On the one hand, some signatures of a crumpled globule state are visible
(particularly visible \textit{post factum}) in simulation work
\cite{Rosa_Everaers_PLOS_2008}. On the other hand, the naive original idea of a crumpled globule \cite{crumpled2,crumpled1} requires much deeper understanding.  The major source of such new understanding over the last few years was computer simulation of both lattice
\cite{Vettorel_Grosberg_Kremer_2009,PhysicsToday_Image} and off-lattice models
\cite{Rosa_Everaers_Looping2010,Melt_of_Rings_Statics_2011,Melt_of_Rings_Dynamics_2011,
Heermann_chromatin_loops,Heermann_polymer_loops_chromatin} (see also
\cite{Linear_contaminants,Comparing_Lattice_and_off-Lattice}). Here we will review the status of our understanding of topologically restricted dense polymers and their relevance to genome folding.

\subsection{Why melt and why rings?}\label{sec_sub:why_melt_why_rings}

To demonstrate this relation, we follow our main guiding theoretical
idea, namely that topological constraints are bound to play a central
role in genome folding and the overall nuclear architecture. This may
seem surprising given that DNA in a eukaryotic cell has open ends.
However true, open ends do not cancel the topological constraints,
because reptation, the leading mechanism of topological relaxation
in linear polymers \cite{pgdg,DoiEdwards}, is likely to be
suppressed for chromatin. The main reason why reptation is not
relevant is because the chromatin fiber is very long. Rosa and
Everaers \cite{Rosa_Everaers_PLOS_2008} estimate that the reptation time
for human chromatin is orders of magnitude longer than
the characteristic times of all known cellular processes.  Thus, once in
a crumpled globule state it would take far too much time to relax
into an entangled equilibrium state. Phenomena like these are also
discussed in the context of polymer rheology
\cite{Vettorel_Developing_Entanglements,McLeish_2007}. Furthermore,
there are at least two additional factors suppressing reptation of
chromatin fibers in the cell nucleus. First, telomere regions at the
end of chromosomes
\cite{Olovnikov,Telomeres1,Telomeres2,Telomeres3}, by virtue of
their peculiar sequences which include huge numbers of repeated
short motives, are likely to form bulges preventing reptation
(similar to the mechanism described by de Gennes \cite{DeGennes_1968}).

Second, some parts of the chromatin fiber, particularly
heterochromatin, are likely to have attachment points to the inner
surface of the nuclear envelope (lamina). Also, topological enzymes are
likely to play only very limited role for the interphase nucleus, if
any (see more details on that below in Section
\ref{sec:qualitative_territories}).

Thus, let us consider the idea that the chromatin fiber cannot cross
itself and does not reptate as a working hypothesis. What are the
consequences of such an assumption, and how do they compare to the
data? The main ingredient of a model must be the large amount
(length) of chromatin fiber stored in a restricted volume at high
concentration with topological restrictions. We also adopt the
hypothesis \cite{crumpled2} that the topological state of chromatin
in the nucleus is very simple. That is, there is either a complete lack
of knots or at most a very few rather simple ones. Although
mathematically knots are defined for closed loops only, the concept
of knots is still reasonable for very long open strings, such as
chromatin.  (We note in passing that the idea of approximately
defined knots in open strings recently gained popularity
in the context of
proteins, e.g., \cite{Millet_Sulkowska_Onuchic_protein_knots}).
Therefore, we argue that the \textit{simplest theoretical model}
which meets all of the above conditions is a \textit{melt} of very
long \textit{unknotted and nonconcatenated} rings. This model
system mimics the idea of chromatin being (nearly) unknotted and of
pretty high concentration in the nucleus.  The obvious drawback of
rings, namely that they have no ends, is assumed to be of marginal
importance for conformations of very long polymers, once they have
been prepared in a non-entangled starting configuration.

The idea that chromatin, because of the lack of reptation, can be
modeled by a system of nonconcatenated rings was first suggested by
Rosa and Everaers \cite{Rosa_Everaers_PLOS_2008}. They also provided
arguments that this approach could explain the phenomenon of
chromosome territories.  In the same work
\cite{Rosa_Everaers_PLOS_2008} they simulated very long chains
with open ends, with the idea that they more faithfully represent
chromatin. The advantage of the ring model is that it allows for clean
conclusions to be drawn regarding the role of topological constraints. Unlike for
chains with open ends, results are not restricted to times which are orders of magnitude below
the equilibration time. This avoids the difficult task of mapping time
scales between real and simulated chromatin. The ring model is particularly
articulate because conformations within the computer simulation time
are fully equilibrated. If the rings display any kind of crumpled globule
behavior, this is a strong indication that the driving force to
tangle for very long open chains is rather weak and probably not
relevant on times typical for interphase chromosomes.

Another, easier question is that of density.  The DNA is packed in
interphase chromosomes at the density of a concentrated solution. In
polymer physics such a system is best described as a melt of
properly defined blobs (see, e.g.,
\cite{pgdg,RedBook,RubinsteinColby}).  Although the precise
definition of such blobs for chromatin is by no means trivial, we
assume that our consideration is performed at the level of blobs,
which can be estimated as being not larger than about \unit[300]{nm}
(see below), as suggested by small angle neutron scattering
experiments \cite{Chromosome_Fractality_Scattering}. Thus, we
consider a melt or a system where the polymers fill essentially all
available space homogeneously. It turns out that such a
melt-of-rings-based approach quite directly yields a natural
explanation of chromosome territories.

\subsection{Physical parameters of chromatin fiber as a polymer}

\subsubsection{Linear density and persistence length}

As we discussed above, chromatin fiber is not perfectly well
defined.  Nevertheless, to approach its polymer physics we must have
some idea about its polymer parameters.  Luckily, the cornerstone
of polymer physics is the concept of universality \cite{pgdg}: not
very many parameters are required to approach the generic
properties, such as those related to entanglements.  In fact, four
parameters are most important: polymer length, persistence length,
density and the resulting entanglement length.  Let us discuss their estimates.

Usually, the genome length $N$ is known in terms of the number of base
pairs (see, e.g., Table \ref{tab:examples_of_organisms}).  That
means, the dsDNA length $L$ is also known, $L = bN$, where $b =
\unit[0.34]{\frac{nm}{bp}}$. In other words, the ``linear density'' of
the double helix is $1/b$ which is roughly 3 base pairs per nanometer.
The similarly defined ``linear density'' for the \unit[10]{nm} fiber is
about 20 base pairs per nanometer.  The linear density $1/B$ for
chromatin fiber is not known exactly, but experimenters believe it
to be above or close to $1/B \approx \unit[40]{bp / nm}$
\cite{Dekker_Chapter_7}. For example, for the human (diploid) genome of
altogether $\unit[6.6 \times 10^9]{bp}$, this linear density gives a
total length of the chromosome fiber of $\unit[6.6 \times 10^9]{bp}
/ \unit[40]{bp / nm} \approx \unit[16]{cm}$.

The persistence length of bare dsDNA is close to \unit[50]{nm},
while for
chromatin fiber it is difficult to determine
accurately. It is certainly longer than that of dsDNA, and the opinions
seem to be converging on a number around $L_p = \unit[150]{nm}$
\cite{Dekker_Chapter_7} or Kuhn segment $l_K= \unit[300]{nm}$. The
important fact for our purposes is that chromatin fiber is a
flexible polymer on the length scales of our interest and thus can
be discussed in terms of the standard theories of dense systems of
flexible or semi-flexible polymers.

Assuming the linear density is known, one can estimate that the chromatin fiber
Kuhn segment consists of about $l_K/B \approx
\unit[12000]{bp}=\unit[12]{kbp}$. Another interesting estimate is
based on the density of chromatin in a human cell nucleus which is about $\rho
\approx \unit[0.015]{bp/nm^3}$ (this number corresponds to
$\unit[6.6 \times 10^{9}]{bp}$ in the volume of a sphere with
diameter $\unit[10]{\mu m}$; it is equivalent to 1\% DNA volume
fraction indicated in Table \ref{tab:examples_of_organisms}, as the
volume of one base pair is very close to $\unit[1]{nm^3}$).  This
corresponds to a number density of Kuhn segments of $\rho_K = \rho
B/l_K \approx \unit[8 \times 10^{-6}]{nm^{-3}}$. The dimensionless overlap
parameter is then $\rho_K l_K^3 \approx 22$, a value not unusual for
regular synthetic polymers as well.  Thus the chromatin pieces
overlap strongly on the scale of the persistence length, which
supports the picture of a melt/dense solution of flexible polymers.

In the usual polymer physics framework, another group of parameters
has to do with volume interactions between polymer segments, and
includes virial coefficients, excluded volume, Flory
$\chi$-parameter, and the like. These parameters are important to
determine whether a polymer chain swells or collapses, etc.  In our
case, the volume behavior of chromatin is controlled by interactions
with proteins, including histones and non-histones, such as cohesins
and condensins \cite{Bloom_Kerry_Review_2010, Cell_2013_Proteins, Cell_2013_Proteins_NV}. Luckily, we do not
have to worry about it, at least to the first approximation, because
we assume to know the overall density of chromatin, as discussed
above. For a polymer system of a given density, its spatial
organization does not depend very much on the mechanism controlling
its density, whether it is a proper combination of attractive and
repulsive volume interactions or confinement by an outside envelope.
We will rely on this approximation. Of course, this is only good for
an averaged description, the simplest version of mean field.  It
does not capture, for instance, the fact that some parts of
chromatin fiber can be methylated or acetylated more than others,
leading to uneven excluded volume parameters and an uneven density
distribution in space. Also, it does not capture active processes,
transcription factories, and a myriad of other activities happening in the
nucleus. Nevertheless, here we adopt the simplest uniform
approximation, which we view as the necessary first step.

\subsubsection{Entanglement length of chromatin}
\label{sec_sub:estimate_entanglement_length}

As we mentioned previously, our goal is to explore the role
of topological constraints. In polymer physics, the known fruitful
way to approach this kind of problem is in terms of entanglements.
The concept of entanglement is in fact quite subtle.  Although it
arises from the simple fact that polymers are not phantoms and two
pieces of (the same or different) polymers cannot pass through one
another, entanglement is fundamentally a many-chain phenomenon. It
is very common when two chains may appear not entangled to one
another, but they are made entangled by a third chain nearby and
so on. Impressively, in the melt of linear chains of large
length $N$ each coil overlaps with $\sim \sqrt{N}$ of other coils,
but they collectively manage to create a much larger number of order
$N$ of entanglements for the given coil. It is a highly non-trivial
result that all these collective effects can be effectively
described by a single parameter, namely the entanglement length
$L_e$. This parameter characterizes the average chain length such that on
smaller length scales topological constraints are unimportant while
they dominate on larger length scales. A good way to think about it
is to imagine that the chain faces an un-crossable obstacle, on
average, once over length $L_e$. Significantly, it may be that the
distance between entanglements is small $L_e < l_K$ or large $L_e >
l_K$ compared to the Kuhn segment. In the former case the chain is nearly
straight between the obstacles while in the latter case it is a nearly
free coil between obstacles. We cannot possibly do justice to
this subject here and refer the reader to textbooks
\cite{pgdg,DoiEdwards,RedBook,RubinsteinColby}.

An estimate of the entanglement length, $L_e$, is more involved than
that of the previously discussed quantities. In principle, the most
direct measurement of $L_e$ is usually based on rheology. Without
having access to rheological data, we will rely on the large body of
knowledge about a variety of polymers, with densities and
flexibilities varying by several orders of magnitude
\cite{Colby1992,Fetters1999,Everaers_Kremer_Science,Primitive_Path_Everaers_2005,Primitive_Path_Everaers_2008}.
Specifically, we will employ the interpolation formula
\be L_e = l_K \left[\left( \frac{1}{c_{\xi} \rho_K l_K^3}\right)^{2/5}+ \left( \frac{1}{c_{\xi}
\rho_K l_K^3}\right)^{2} \right] \ , \label{eq:Le} \ee
where $c_{\xi} = 0.06$ \cite{Primitive_Path_Everaers_2008} fits a
huge variety of systems ranging from fully flexible polymers to
rather rigid semi flexible systems.  Given the uncertainty in our
knowledge regarding the linear density, and given also that
different cells may have nuclei of different volumes, we can only
indicate a range of possible values of $L_e$, as shown in Table
\ref{tab:entanglement_length_estimates}.

\begin{table}
\caption{Estimates of entanglement length for chromatin fiber. The
value of the linear density for chromatin fiber is not known very
well, while the nucleus diameter can be different from cell to cell.
Accordingly, we present the whole spectrum of entanglement length
estimates. In all cases the calculations are done using formula
(\ref{eq:Le}) for entanglement length $L_e$, while the number of
base pairs between entanglements is calculated as $N_e = L_e/B$. The
case $1/B = \unit[120]{bp/nm}$ and $D = \unit[10]{\mu m}$ was used
by Rosa and Everaers \cite{Rosa_Everaers_PLOS_2008}. The Kuhn
segment was assumed to be $l_K = \unit[300]{nm}$.
\label{tab:entanglement_length_estimates}}
\begin{tabular}{||l||c|c|c||}
\hhline{|t:=:t:=t==:t|}

Linear density & $D=\unit[5]{\mu m}$ & $D=\unit[10]{\mu m}$ & $D=\unit[15]{\mu m}$ \\

\hhline{||-||---||} $\frac{1}{B} = \unit[40]{bp/nm}$ & $L_e = \unit[0.1]{\mu m}$ & $L_e = \unit[0.35]{ \mu m}$ & $L_e = \unit[1.6]{\mu m}$ \\

& $N_e=\unit[4 ]{kbp}$ & $N_e=\unit[14]{kbp}$ & $N_e=\unit[64]{kbp}$ \\

\hhline{||-||---||} $\frac{1}{B} = \unit[80]{bp/nm}$ & $L_e = \unit[0.15]{\mu m}$ & $L_e = \unit[0.7]{\mu m}$ & $L_e = \unit[5]{\mu m}$ \\

& $N_e=\unit[12]{kbp}$ & $N_e=\unit[56]{kbp}$ & $N_e=\unit[400]{kbp}$ \\

\hhline{||-||---||} $\frac{1}{B} = \unit[120]{bp/nm}$ & $L_e = \unit[0.2]{\mu m}$ & $L_e = \unit[1.3]{\mu m}$ & $L_e = \unit[11]{\mu m}$ \\

& $N_e=\unit[24]{kbp}$ & $N_e=\unit[156]{kbp}$ & $N_e=\unit[1320]{kbp}$ \\

\hhline{|b:=:b:=b==:b|}

\end{tabular}
\end{table}

From our knowledge of synthetic polymers, we tend to expect a realistic value of $L_e$ in
between the extremes indicated in Table \ref{tab:entanglement_length_estimates}, perhaps somewhat
closer to the lower end, around $\unit[300]{ nm}$. In summary, one can state that the entanglement length
$L_e$ of chromatin fibers at conditions typical for a human cell nucleus would most probably be
below $\unit[500]{nm}$. This holds for most organisms with diploid cells (cf. Table 1). Compared
to basic results from polymer theory these systems are deeply in the regime where topological
effects dominate all relevant structural and relaxation processes.

In this context it is interesting to compare the above estimates to
nuclei of other cell types, particularly that of yeast. Not only
is the total length of the genome much smaller, more
importantly, the density in the nucleus is about a factor of three
smaller, cf. Table \ref{tab:examples_of_organisms} (while for
chicken nuclei it is larger). It has a total length of $1.2\times
10^7$ base pairs in altogether 32 chromosomes. On average each
chromatin fiber thus contains only 137,000 base pairs. Assuming the
same chromatin fiber structure for yeast as for human diploid cells,
which probably will only be very approximate, one arrives at about
$\unit[40]{\mu m}$ total contour length divided among 32 chromatin
fibers. Considering the reduced density, the average total contour
length per fiber is only a few times more than or even close to
$L_e$. Thus for yeast, topology effects are expected to be very weak.

\subsection{A physical analogy: rheology of non-entangled melts}
\label{sec_sub:rheology}

It is instructive to relate the above analysis to known results from polymer chain dynamics and
rheology. Rosa and Everaers \cite{Rosa_Everaers_PLOS_2008} have argued that the dynamics of
chromosomes in the cell nucleus are orders of magnitude slower than all known biological processes.
Based on the reptation concept, they estimate the overall relaxation
time in such a system to be well above the average life time of a human being. In contrast the
overall relaxation time of a melt of nonconcatenated rings is much smaller than that of an
entangled melt of linear chains, as we will see below \cite{Melt_of_Rings_Dynamics_2011}.
Considering the stability of
chromosome territories, the specific question is whether there is a strong tendency to entangle,
which then would need specific measures to be prohibited, or whether such a tendency does not
exist.

As already mentioned, one expects a pressure difference $\Delta p$ in a computer simulation
between a melt of nonconcatenated rings and a melt of linear polymers of the same chain length
and density. This is because of the different number of degrees of freedom and of the topological
constraints. $\Delta p$, however, has not been detected so far. Though too small to be detected,
the total free energy increase per chain compared to a melt, roughly $k_BTb^3 \Delta p N$ with $N$
being the number of repeat units per chain, can be quite large. Switching off the topological
constraints would lead to a very fast relaxation of the polymer globules towards Gaussian
conformations through an isotropic expansion of the polymers. Indeed for linear chains of length
up to several $L_e$ this holds true
\cite{Sadler_1991,DeGennes_1995,Vettorel_Developing_Entanglements}, even though chain
connectivity and non-crossability are fully maintained. This rapid relaxation does not occur,
when the chains become much longer. Rastogi et al. \cite{Rastogi_2005} studied the processing
properties of a polymer melt, which was created upon melting a dense agglomeration of polymer
crystallites, each of them containing only one single, but very long chain. They observed a
significant reduction in the apparent, transient viscosity compared to an entangled melt of
otherwise identical properties. Indeed, studying the relaxation of a melt of compact globules
into an entangled melt revealed a rapid expansion for short chains and a very slow process for
very long chains ($b N \geq \cal{O}$$ (50 L_e)$)
\cite{Vettorel_Developing_Entanglements}.
This phenomenon can be understood in terms of the elastic distortion due to
``non-cooperative'' reptation \cite{McLeish_2007}. That is, the reeling out of chain ends into the
surrounding chains creates an elastic deformation of globules, making the relaxation process very
slow. For nonconcatenated rings that would require reeling out of doubly-folded pieces, which are
subject to an additional entropy penalty. Unfortunately there is no systematic study of this
phenomenon so far. Despite this, these results indicate that for the diploid cells of higher organisms
no complicated biochemical apparatus is required to keep the chromosomes segregated on the
biologically relevant time scales once they are ``prepared'' in a territorial arrangement.

\section{Melt of rings and related polymer models}
\label{sec:melt_of_rings_and_polymer_models}

\subsection{Model and question formulation}
\label{sec_sub:model_and_question_formulation}

Understanding the conformational geometry and statistics of nonconcatenated rings in the melt
turned out to be an unexpectedly difficult
challenge \cite{Won_Bo_Lee_Melt_Unconstrained_Rings}. To emphasize the generic character of the problem, even beyond the previous discussion, and
its independence of any chemical or microscopic details, it is useful
to start with a purely mathematical formulation.

Imagine a piece of cubic lattice in space with number of nodes $K \gg 1$ occupied by $M$
closed loops or rings of $N$ steps each, leading to $NM=K$. In this case, space is completely filled which corresponds to the melt assumption where each blob occupies a node giving a volume fraction $\phi=1$.  Importantly, we assume
that each ring is unknotted and rings are not concatenated. Furthermore, no node is occupied more than once.

This is the description of the simplest model of a concentrated set of unknotted and nonconcatenated rings, the subject of our attention. Of course, this system can be considered off-lattice, and can be equipped with further details. The main issue, however, is the interplay between  conformations and packing of homogeneously space-filling polymers, i.e., $\phi \simeq 1$, and topological constraints of being unknotted and nonconcatenated.

Here is a list of simple questions with regard to the statistical
properties of this model, and which are most closely related to the experimental data being collected by methods such as FISH and 3C:
\begin{itemize}
\item How does the spatial extension (e.g., averaged gyration radius) of a ring squeezed between others depend on the ring length $N$?  We expect the dependence to be a power law:
\be R_g^2 \sim N^{2\nu} \ .  \label{eq:definition_of_nu} \ee
The index $\nu$ defines whether rings can form territories ($\nu
= 1/3$ for spatial dimension $d=3$) or not ($\nu > 1/3$).
\item How does the size of the subchain (e.g., its gyration radius or end-to-end Euclidean distance) $r(s)$ depend on the subchain arc length $s$. We expect it to be governed by the same index $\nu$ at sufficiently large $s$, so that $r(s) \sim s^{\nu}$.
\item  What is the probability $P(s)$ that two monomers separated by arc length $s$ will be in contact in space?  Again we expect this contact probability, or loop factor, to follow a power law in $s$, with a power $\gamma$, where the relation to $\nu$ is at least not clear:
\be P(s) \sim s^{-\gamma} \ . \label{eq:definition_of_gamma} \ee
\item How many monomers of one ring are in contact with monomers of other rings?  How many monomers of a subchain are in contact with other subchains?  We expect this fraction of a ring, which could be called ``surface'' or easily accessible fraction, to be governed by another exponent $\beta$:
\be n_{\mathrm{surf}} \sim N^{\beta} \ . \label{eq:definition_of_beta} \ee
\end{itemize}

There is also a separate set of questions regarding the dynamics of rings in concentrated systems (see Section \ref{sec:Dynamics} below), but for now we concentrate on the equilibrium statistics in relation to chromatin experiments.

\subsection{Theoretical approaches}\label{sec_sub:theoretical_approaches}

\subsubsection{Flory type theories.}

A significant step was the work of Cates and Deutsch \cite{Cates_Deutsch_1986}, who considered a melt of nonconcatenated rings and arrived at the prediction that the size of a ring scales as
\begin{equation}\label{eq:Cates_Deutsch_result} R \sim b N^{\nu} \ , \ \ \ \mathrm{where} \ \ \ \nu = 2/5 \ . \end{equation}
Their approach follows a classical Flory theory of regular polymers with excluded volume and makes the argument that the equilibrium size of the
ring is determined by the balance of two factors: entropy loss of the ring itself when it gets more compact than its preferred Gaussian size, and entropy loss due to the topological constraints with the surrounding rings if the ring size increases by protruding loop-like ``tentacles''. The first factor is estimated as $Nb^2/R^2$, where $N$ is the number of monomers in the ring, while $R$ is the ring size to be determined, and $b$ is a microscopic length scale such as the Kuhn segment. The second contribution to the entropy is estimated as follows: if every ring has size of order $R$ and, therefore, lives in a volume $R^3$, while its own volume is only about $\sim Nb^3$ (assume for simplicity that the chain thickness is governed by the same length scale $b$), then about $R^3/Nb^3$ different rings cohabit one and the same volume. If we assume that every one of these rings forces our chosen ring to loose entropy of order unity, then we arrive at the following estimate of the overall free energy:
\begin{equation}\label{eq:Free_Energy_Cates_Deutsch}
    \frac{F}{k_BT} \sim \frac{Nb^2}{R^2} + \frac{R^3}{Nb^3} \ ,
\end{equation}
where $k_BT$ is temperature in energy units.  Minimization with respect to $R$ yields the result (\ref{eq:Cates_Deutsch_result}). Surely, this argument is very much open to criticism, much more so than the similarly looking classical Flory theory of a coil in good solvent. For instance, the first term assumes the rings to be Gaussian and neglects that the ring itself is unknotted. In reality, the free unknotted ring is swollen \cite{RingsNotGaussian}, thus its statistics are not Gaussian for topological reasons. Therefore, entropy loss due to compression is larger and grows faster with decreasing $R$. The second term in Eq. (\ref{eq:Free_Energy_Cates_Deutsch}) is even more problematic. It is not clear at all why every one of the cohabiting rings produces an entropy loss of order unity.  One can try to improve this estimate by assuming, for instance, that the osmotic pressure of surrounding rings is a many-body type of phenomenon such that the corresponding free energy scales as $(R^3/N b^3)^{\alpha}$ with 
some properly chosen power $\alpha$ (see \cite{Cates_Deutsch_1986}). Instead of formula (\ref{eq:Cates_Deutsch_result}) and index $2/5$, such an approach yields index $\nu = (1+\alpha)/(2 + 3 \alpha)$, which can be anything in the
interval $2/5 > \nu > 1/3$ depending on $\alpha$. Of course, this only emphasizes the physically unjustified character of the estimate. Nevertheless, the very idea of a proper balance of entropy losses due to intra- and inter-ring contributions behind Eq. (\ref{eq:Free_Energy_Cates_Deutsch}) deserves serious attention as a basis for possible future improvements.

One noteworthy attempt of such improvement was undertaken by T. Sakaue
\cite{Sakaue_Rings_Brief_PhysRevLett.106.167802,Sakaue_Rings_Detailed_PhysRevE.85.021806}. It is also based on minimization of free energy for a test ring, qualitatively similar to Eq. (\ref{eq:Free_Energy_Cates_Deutsch}) in the sense that there is one term which favors compression because the test ring looses entropy by making long loopy protrusions among surrounding rings, and there is another term which disfavors compression because the loss of entropy of the test ring itself. In such a general form, the idea is undoubtedly correct. Such concepts are well known, as for instance discussed for the case of gels in \cite{Grosberg_TwoTypes}. Both terms were estimated in
\cite{Sakaue_Rings_Brief_PhysRevLett.106.167802,Sakaue_Rings_Detailed_PhysRevE.85.021806} in a more involved way compared to \cite{Cates_Deutsch_1986}, but the estimates rely upon some rather \textit{ad hoc} assumptions. In the end, the suggested variational free energy reads
\begin{equation}\label{eq:Free_Energy_Sakaue}
    \frac{F}{k_BT} \sim \ \ln \left( 1 - \frac{R^3}{b^3N \sqrt{\tilde{N}_e}} \right) + \frac{N^3b^6}{R^6} \ ,
\end{equation}
where numerical coefficients of order unity are all dropped out, but the entanglement length $\tilde{N}_e$ is defined in a peculiar non-standard way, different from the usual definition by a factor of about $2$ or more. This suggested form of the free energy automatically requires that very long rings in a melt are compact, because $R$ asymptotically is growing faster than $N^{1/3}$, making the argument in the log eventually negative. However, if we assume $\tilde{N}_e \gg 1$, then on the way to the final asymptotics $R \sim
N^{1/3} \tilde{N}_e^{1/6}$, valid for $N \gg \tilde{N}_e$, the free energy
(\ref{eq:Free_Energy_Sakaue}) predicts a region of intermediate asymptotics $R \sim N^{4/9} \tilde{N}_e^{1/18}$ valid at $\tilde{N}_e \gg N \gg 1$ (the latter result is not mentioned in Refs.
\cite{Sakaue_Rings_Brief_PhysRevLett.106.167802,Sakaue_Rings_Detailed_PhysRevE.85.021806}, but can be derived from the free energy (\ref{eq:Free_Energy_Sakaue})). Although this theory involves many poorly justified assumptions, it does capture the existence of a relatively wide cross-over region, with seeming power $4/9$ which is close to the $2/5$ result of Cates and Deutsch (\ref{eq:Cates_Deutsch_result}).

An important aspect of free energy estimates, such as Eq.
(\ref{eq:Free_Energy_Cates_Deutsch}) or
(\ref{eq:Free_Energy_Sakaue}) is that they represent a small, but
decisively important, correction to the very large contribution
from repulsive forces of interaction between monomers at high
density. These corrections are decisive because they depend on $R$
and their structure is determined by the topological constraints,
such as absence of knots in any particular ring and absence of links
for any group of rings. For instance, there is no doubt that the
osmotic pressure of a concentrated system of nonconcatenated rings
is larger than that of a similarly concentrated system of linear
chains (for large $N$ and at the same density).  However, this
correction is so small, compared at the dominant term produced by
excluded volume interactions between monomers, that none of the
presently available simulations is sensitive enough to detect it, as
previously mentioned.

\subsubsection{Crumpled, or fractal, globule}

Another possible idea regarding rings in the melt is based on
the convolution of three hypotheses. The first hypothesis assumes that the
fast collapse of a single polymer chain upon abrupt solvent quality
quench produces a peculiar state, called crumpled globule
\cite{crumpled1}. This state is dense, self-similar, and free of
knots and should have $\beta=2/3$. The second (least proven)
hypothesis is that this state of a crumpled globule is equilibrium
for any isolated chain collapsed in a poor solvent with the constraint
that knots are excluded, for instance, for an unknotted ring. The
third hypothesis suggests that a ring squeezed by its neighbors in
the melt and a ring collapsed in a poor solvent should be similar and,
therefore, both should be crumpled globules. This latter hypothesis
will be especially questioned by the simulations. Thus, the idea is
that a crumpled globule results when the polymer is
forced to adopt a compact conformation without knots, whether the
lack of knots is an imposed topological condition, as in a ring, or
it results from slowness of reptation, as in rapid collapse (where
collapse should be viewed as ``rapid'' if it occurs faster than the
relevant relaxation time scale; in practice, it may still be rather
slow by conventional standards).

According to arguments developed in \cite{crumpled1}, the signature
of a crumpled globule is that every subchain of sufficient length
$s$ is collapsed in itself, and has the size of order $\sim b
s^{1/3}$. In other words,
\be r(s) \sim s^{\nu} \ \mathrm{with} \ \nu = 1/3 \ , \label{eq:crumpled_prediction}
\ee
which is assumed to be true at $s > s_{\textrm{min}}$ with $s_{\textrm{min}} \sim
N_e$. That also means the chain itself, on the scale beyond the
entanglement length $N_e$, has fractal dimension $3$. The idea of
justification goes back to the well established fact of topological
repulsion between nonconcatenated rings. For instance, the second
virial coefficient of two nonconcatenated rings in dilute solution
is close to $R_g^3$ and is practically independent, at least in the
scaling sense, of the real excluded volume of monomers
\cite{Maxim_second} (see also recent work
\cite{Bohn_Heerman_Topological_Loop_Repulsion} and references
therein). Nevertheless, the concept the crumpled globule remains a
hypothesis.

Furthermore, we should emphasize that all three aspects mentioned above need careful attention:
although the simulations in \cite{HiC_Science_2009} do confirm the formation of a crumpled globule upon
fast chain collapse, it is not proven that this scenario is valid for all mechanisms of squeezing
(e.g., by external field, by poor solvent, etc). It is likely but not proven that an equilibrium
collapsed ring is a fractal. Additionally, it is not clear that rings squeezed in the melt form the same type
of fractal as single collapsed rings, simply because the surface of a collapsed globule is very
rough in the former case while smooth and dominated by surface tension in the latter. The most likely
possibility is that all of these systems are crumpled globules of somewhat different kinds, but
this remains to be properly understood. An attempt to understand the nature of crumpled globules
more deeply faces serious mathematical challenges
\cite{NechaevVasilyev_2005,NechaevVershik_BrownianBridge}.

A direct computational test of the hypothesis that the typical
conformation of a collapsed chain without knots is a crumpled
fractal globule was attempted in the work
\cite{Lua_Borovinskiy_Grosberg_POLYMER_2004}. For that purpose,
Hamiltonian walks (i.e., lattice conformations which fill every site
of a given lattice segment once and only once) were generated on
cubic segments of a cubic lattice up to the size of $14 \times 14
\times 14$. A rather clear tendency toward segregation of subchains
was observed, but the length of the polymer was insufficient to arrive
at conclusive results.

The question of the long-term stability of crumpled structures was recently addressed by simulation \cite{Schiessel_2013}. For the particular crumpled conformations which were used as initial conditions, long-term stability was not observed.

Interestingly, conformations that are similar to crumpled are implicated in certain scenarios of active (ATP dependent) formation of chromatin structures, by so-called loop-extruding enzymes \cite{Marko_Loop_extruding}.

\subsubsection{Lattice animal model}\label{sec:Lattice_animal_model}

Although the intent of the above approaches is to describe the
restriction of accessible conformations due to the topological
constraints, the topology is present there only indirectly. In
general, theoretical ideas about polymer topology revolve around
the idea of an effective tube suggested originally for networks
\cite{Edwards_first_Tube} and most widely known in the context of
reptation
\cite{pgdg,DoiEdwards,RubinsteinColby,RedBook,GiantMolecules}. Tubes
and reptation ideas, unfortunately, have no direct applicability for
nonconcatenated rings.  Since the tube is a sort of ``topological
mean field'', a similar idea for the melt of rings is a ring squeezed into a  lattice of obstacles, just like in the very first work of Edwards
\cite{Edwards_first_Tube}. That means, we should imagine placing one
ring into a lattice of immobile, un-crossable, and infinitely long
straight spikes such that the ring is not tangled to the lattice,
i.e., it must be topologically possible to take the ring out of the
lattice. This model was examined in the works
\cite{Khokhlov_Nechaev_1985,Rubinstein_PRL_1986,Nechaev_Semenov_Koleva_1987,Obukhov_Rubinstein_Duke_PRL_1994}.
The conformation of a ring in such a lattice is well understood, it
represents a so-called lattice animal, i.e., the ring double-folds
along an annealed branched structure. The description of this system
is mathematically beautiful \cite{NechaevVasilyev_2005}, and it is
undoubtedly correct for a single ring in the lattice of obstacles.

The applicability of a lattice of obstacles and annealed branched structure in the self-consistent situation of many rings is not obvious.  It is a difficult computational task, so far unresolved, to check for such structures in the computationally generated conformations of rings. The difficulty arises from the fact that branches, if they are real, exist only on a scale significantly larger than $N_e$.  This motivates the idea to assume the existence of such an annealed branched structure, examine the consequences of the assumptions, and compare them with computational data. Such an attempt was undertaken by one of us (AYG) in the recent work \cite{Generalized_Lattice_Animal}, and the results are rather encouraging.  First, this logic yields a rather straightforward proof that $\nu = 1/3$, i.e., every ring collapses onto itself and different rings form territories. Interestingly, the backbone of a branched lattice animal inside its territory behaves like a self-avoiding random walk. That means the backbone 
length $L$ is estimated from the condition $L^{3/5} \sim N^{1/3}$ (ignoring here $N_e$; see \cite{Generalized_Lattice_Animal} for details).  This result may seem counterintuitive because the backbone is a polymer and normally self avoidance is screened under melt conditions \cite{pgdg,DoiEdwards,RubinsteinColby,RedBook}.  The fact of the matter is that the screening effect in our case is exactly compensated by the unusually large excluded volume, as it is controlled by the side chains of the branched structure. And the distribution of material between the backbone and the side branches in the annealed system is self-consistently determined so as to establish the self-avoiding statistics.

These ideas also shed light on other critical exponents for the system, such as the surface exponent $\beta$ and the contact exponent $\gamma$. It turns out that this $\gamma$ is related to the ``other $\gamma$'', which we denote here $\gamma_{\mathrm{usual}}$, and which controls the number of conformations for a polymer \cite{pgdg}.  This relation explains why our index $\gamma$ cannot be computed by any analog of Flory theory, and, more generally, why finding this index is a very difficult task.

The lattice animal model was also used in the work \cite{Arya_PRE} to model chromosomes directly on a phenomenological level, and we return to this later in Section \ref{sec_sub:simulation_of_other_models}.

This completes our discussion of the rather unsettled situation of theoretical concepts aimed at describing the overall size of the rings, or the index $\nu$ (\ref{eq:definition_of_nu}).  Apart from the very recent attempt \cite{Generalized_Lattice_Animal}, no theory we are aware of goes beyond $\nu$ to address contact probabilities, subchain surfaces, and other more detailed conformational properties which are
described by an independent set of exponents which we call $\beta$ and $\gamma$, as defined in Eqs. (\ref{eq:definition_of_gamma}) and (\ref{eq:definition_of_beta}).

\subsection{Peano type space-filling curves and mathematical hypothesis}\label{sec_sub:Peano}

\begin{figure}
  \includegraphics[width=0.45\textwidth]{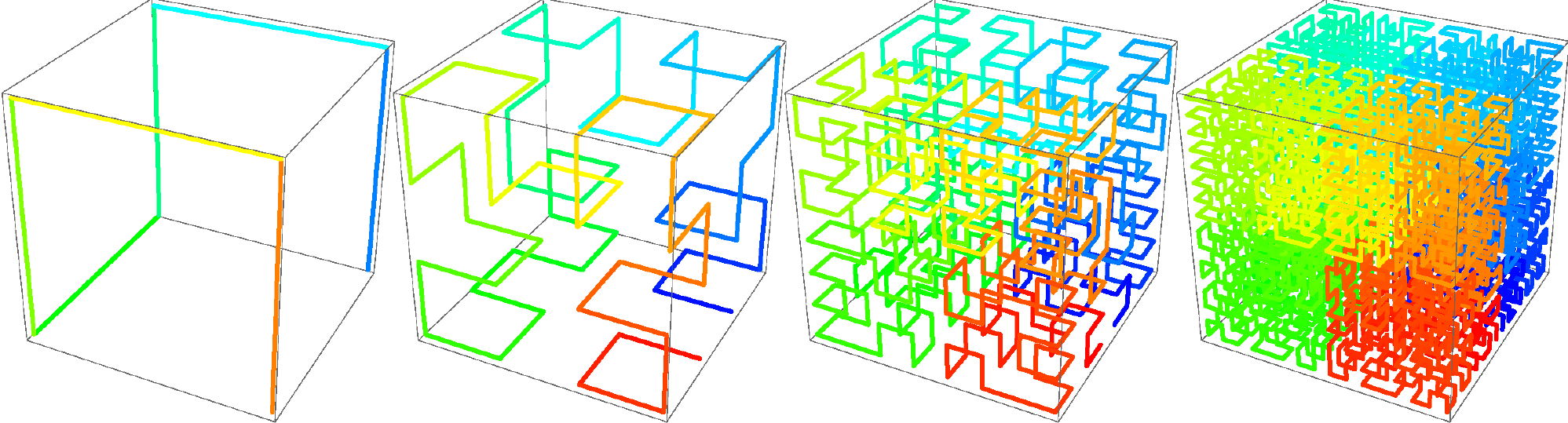}\\
  \caption{Closed Hilbert space-filling curves, called Moore curves,
shown in the first four steps of generation. Both ends are next to
each other on the lattice, which is why the curve can be considered
as a closed loop. The polymer in every case is colored in the
rainbow order from one end to the other. This allows one to clearly
see the ‘territorial segregation’ between subchains.}\label{fig:Hilbert_curve}
\end{figure}

Good realizations of crumpled globule conformations are well known
in mathematics as space-filling curves.  The most widely known
example is the Peano curve in 2D. Other examples include the Hilbert
curve and its closed loop (with two ends connected) version called
Moore curve (shown in Fig. \ref{fig:Hilbert_curve}), as well as
Sierpinski, Lebesgue, and Gosper curves.  For a modern review of
this topic, including computer algorithms to generate these curves,
see the book \cite{Space_Filling_Curves_Book}. See also
\cite{Use_of_space-filling_curves} for some more recent unexpected
applications of space-filling curves in data sorting and
parallelization of computer algorithms.

Space-filling curves are usually constructed via recursive
algorithms, and, in the limit, they are true self-similar fractals.
Of course, unlike in the mathematical literature where the nature of the limit is the central aspect, for our purposes we only need a high, but
finite level of iteration. In this sense, for instance, the 3D Hilbert
or Moore curves in Fig. \ref{fig:Hilbert_curve} represent true
fractal globules: they are unknotted and self-similar by construction.

\begin{figure}
  \includegraphics[width=\columnwidth]{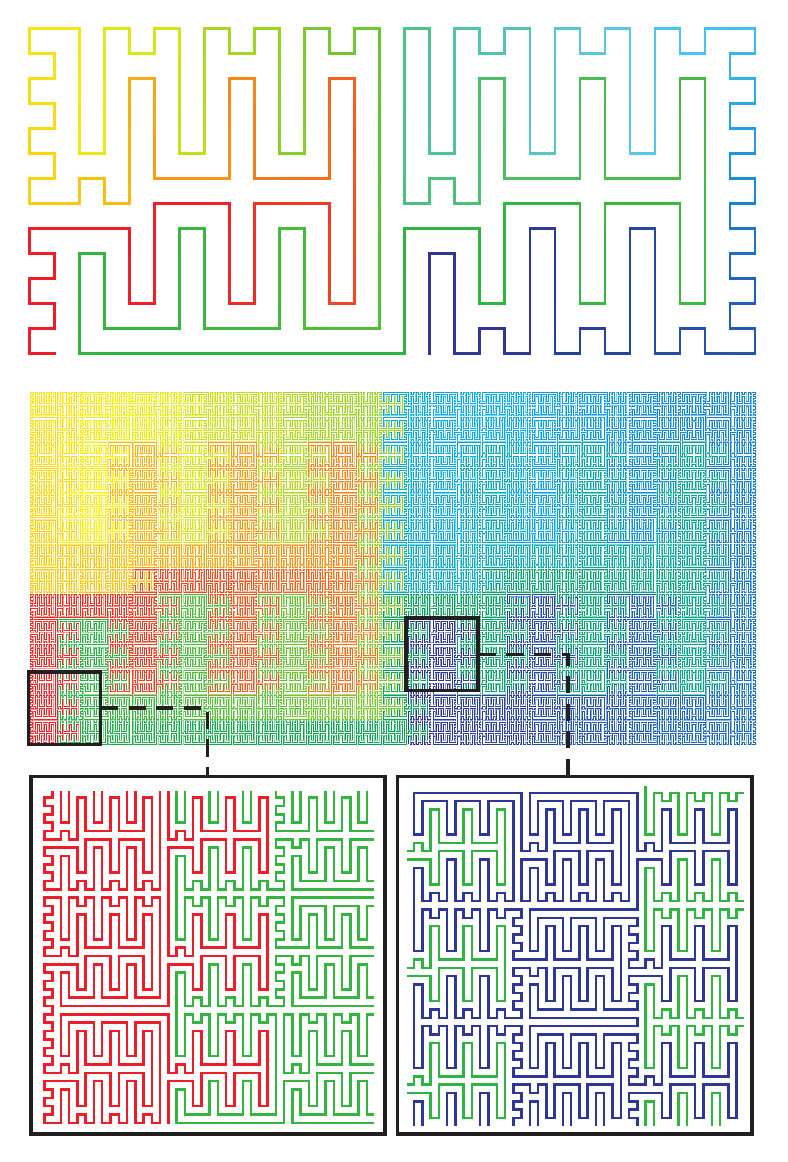}\\
  \caption{A 2D example of a space-filling curve with large fractal
dimension of inter-territorial boundaries. First (top) and second
(center) steps of generation of the curve. Two magnified sections of
the second step are shown at the bottom. Coloring of the polymer in
rainbow order from one end to the other enables one to see
‘territorial segregation’ and the rough surfaces between subchains.}\label{fig:Jan_Smrek_curve}
\end{figure}

Classical curves, such as Peano, Hilbert, Moore and Sierpinski space-filling curves, are characterized by pretty smooth surfaces between neighboring folds, which corresponds to $\beta = 2/3$ in 3D
(or more generally $\beta = (d-1)/d$ in $d$-dimensional space). As
we will show below, this yields $\gamma = 4/3$ (or $\gamma =
(d+1)/d$). In the context of chromatin models, the natural question
is this: is smoothness of surfaces, expressed by the value $\beta =
2/3$, an inherent property of regular space-filling curves?
Recently, two of us (JS and AYG) answered this question by
explicitly constructing the entire novel family of fractal unknotted
space filling curves with very wiggly surfaces
\cite{Jan_Smrek_curve}. A 2D example of such a curve is shown in
Fig. \ref{fig:Jan_Smrek_curve}.  In fact, there are curves with
index $\beta$ arbitrarily close to unity from below (and
consequently $\gamma$ arbitrarily close to unity from above).
E. Lieberman-Aiden mentioned to one of us (AYG) a similar result
of his own \cite{Erez_Unpubished_construction}.

Of course, the existence of one or a few fractal space-filling
curves with various surface roughness (values of $\beta$) does not
mean that there are enough conformations of this type to provide
sufficient entropy and to make the state thermodynamically
competitive. In this sense the crumpled globule as a thermodynamic
(macro)state is still a hypothesis, and the entropically
(or probabilistically) dominant values of $\beta$ and $\gamma$ for a
randomly chosen conformation are still not known.

\subsection{Simulation data for the melt of rings}
\label{sec_sub:simulation_data_rings}

The most detailed simulation results on the melt of unknotted nonconcatenated rings are presented in the works \cite{Vettorel_Grosberg_Kremer_2009} using the Monte Carlo method for a lattice model and in back-to-back papers
\cite{Melt_of_Rings_Statics_2011,Melt_of_Rings_Dynamics_2011} using
off-lattice molecular dynamics.  These data along with other available simulation results \cite{Muller96,Brown98b,Suzuki09,Yoon2010} are summarized in Ref. \cite{Linear_contaminants,Comparing_Lattice_and_off-Lattice}. We do not attempt to repeat here all of the many results, but the main outline can be formulated as follows.

Much effort went into the proper equilibration of the samples, and convincing evidence was accumulated to claim that equilibration was achieved.  For the lattice model, rings up to $N=5000$ were considered ($N=10000$ were not fully equilibrated) whereas for the off-lattice systems it was done for rings up to
$N=1600$. In the former case, the entanglement length $N_e$
(\cite{pgdg,DoiEdwards,RubinsteinColby,RedBook,GiantMolecules}) was found to be $N_e \simeq 175$ versus $N_e \simeq 28$ in the latter case. Therefore, in terms of the number of entanglements per chain the achieved lengths were roughly comparable: $N/N_e$ up to about $30$ for lattice systems (and $60$
incompletely equilibrated) and up to about $57$ for the off-lattice case.

In both the lattice and off-lattice systems, the overall ring size,
measured by either $R_e$, the distance between beads $N/2$ apart
along the ring, or $R_g$, approaches with increasing $N$ an
asymptotic behavior which seems consistent with $R_g \sim N^{1/3}$.
At the same time, the dependence $R_g(N)$ exhibits an unexpectedly
wide cross-over region between Gaussian behavior $R_g \sim N^{1/2}$
at the distance smaller than the entanglement length $N < N_e$ and
asymptotic behavior at very large $N$.  In this cross-over region
one expects the intermediate asymptotics close to the estimate of Eq.
(\ref{eq:Cates_Deutsch_result}) $R_g \sim N^{2/5}$ or $N^{4/9}$, cf.
Eq. (\ref{eq:Free_Energy_Sakaue}).

The comparison between different simulation models in terms of
looking at their number of entanglements, $N/N_e$, was justified
for linear polymers in 
\cite{Everaers_Kremer_Science,Primitive_Path_Everaers_2005,Primitive_Path_Everaers_2008}.
This approach appears fruitful also for the rings.  In a recent
paper \cite{Linear_contaminants} it was shown how the data of many
very different simulations collapse on a single
master curve when plotted as $\left< R_g^2(N) \right> \left/\left<
R_g^2(3N_e) \right> \right.$ against $N/3N_e$ (the factor $3$ was
introduced for historical reasons). This universal master dependence
is observed over a very wide range: $0.1 < N/N_e < 150$. It is this
dependence that exhibits a very wide cross-over and eventually
approaches the $N^{1/3}$ scaling.

When this paper was already written, a new simulation work appeared \cite{Rosa_Everaers_Branched_simulation}. In that article, the molecular dynamics simulation results for the melt of unconcatenated rings is generally in very good agreement with our simulations \cite{Melt_of_Rings_Statics_2011}. Additionally, the authors show that simulation results for the melt of rings are in quantitative agreement with more efficient simulations of lattice systems of annealed branched objects coarse grained above the $N_e$ scale. This gives more credence to the idea of the annealed lattice animal representation, discussed above (section \ref{sec:Lattice_animal_model}). This is also highly promising in terms of simulating the larger systems.

Ostensibly an $N^{1/3}$ scaling indicates that different rings are
segregated from one another.  A closer look at the data indicates that
despite the scaling of a compact object $R_g \sim N^{1/3}$ for very
large $N$, the rings do not look at all like smooth rounded globules. By
contrast their shapes are very irregular, and their surfaces are
very rough. In what follows we will pay major attention to the
characteristics of this roughness and the attempts to relate it to
the properties of chromatin.

\subsection{Simulation of other models with topological constraints}
\label{sec_sub:simulation_of_other_models}

In the context of chromatin modeling, rings are only a tool.  Their advantage is that the model
is very cleanly formulated, allowing us to achieve very solid unambiguous results.  The
alternative approach is to simulate very long chains with open ends, following the logic that
real chromatin fibers appear to have their ends available. Two groups followed this
approach
\cite{Rosa_Everaers_PLOS_2008,Everaers_2010_Gene_Colocalization,Rosa_Everaers_Looping2010}, as we
also discuss below.  They prepared initial conformations of long linear worm-like chains in the
form of a pretty densely packed and definitely un-knotted zig-zag conformation, confined to a
thick cylinder. The overall shape was chosen to approximately match the shape of specific
chromosomes (human chromosome 4 and drosophila chromosome 2L) in their condensed metaphase state.
Then they placed the initial polymers into a box with periodic boundary conditions and followed
the slow relaxation of the conformations. They intentionally examined only times shorter than the
overall relaxation of the whole system, under the presumption that time scales were matched to
biological ones and, therefore, one should not worry what happens on the time scale longer than
the cell cycle. More recently they introduced defects or kinks into their worm-like chain model
in order to better reproduce the observed contact probability behavior
\cite{Rosa_Everaers_Looping2010}.

A rather different type of simulation was reported in the work
\cite{HiC_Science_2009} (particularly in its supplemental material)
alongside experimental data. In that work, authors started from a
coil conformation of a discrete worm-like chain and then forced it to
collapse rapidly under the action of a steadily narrowing
``potential well'' acting on every monomer a distance $r$ from the mass
center of the chain as $\phi(r) = k_B T \exp \left[ R_0 \left( r -
R_0 \right)/6 \right]$. $R_0$ was adjusted at every Monte Carlo step
to $0.7 R_{\mathrm{max}}$, where $R_{\mathrm{max}}$ is the
instantaneous maximal distance from the center of mass to any of the
monomers. This strong confinement leads to a very fast collapse.
Once the desired density was reached, $R_0$ was fixed and a standard
MC simulation of a freely jointed chain with excluded volume was
performed. The resulting conformation appeared to be a beautiful
fractal globule with very well pronounced territorial segregation
of different parts and virtually no knots despite open ends.
The contact probability was found to decay as $1/s$, which is very close
to that
observed experimentally. Notice that this algorithm forces the chain to
collapse on a time scale significantly faster than the conformational
relaxation. One of the truly interesting results of these kind of
simulations, compared to those with rings where the conformations
are fully equilibrated, is that the conformations are rather
similar, although maybe not identical.

A more recent simulation study using the same model appears to indicate an
unexpectedly speedy relaxation of this fractal structure.
The interpretation of these findings and their association with
chain length in relation to powers of $N_e$ and other relevant
factors will have to wait until after the details of the work are
made available.

Yet another model was simulated in the work \cite{Arya_PRE}.  It
does not involve topological constraints explicitly, instead the authors
simulated an annealed branched structure, similar to that described
earlier for rings in Section \ref{sec:Lattice_animal_model} and
in the works
\cite{Rubinstein_talk_in_Leiden,Generalized_Lattice_Animal,Generalized_Lattice_Animal_Dynamics}.
As in \cite{HiC_Science_2009}, the model \cite{Arya_PRE} involves
chain compression in real space, but an additional \textit{ad
hoc} assumption is a similarly strong compression of the
``generation number'' (number of branches).

\section{``Territorial polymers'' and chromosome territories: qualitative aspect}
\label{sec:qualitative_territories}

\begin{figure}
  \centering
  \includegraphics[width=0.7\columnwidth]{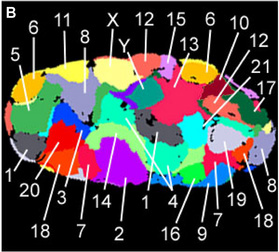}\\
  \includegraphics[width=0.7\columnwidth]{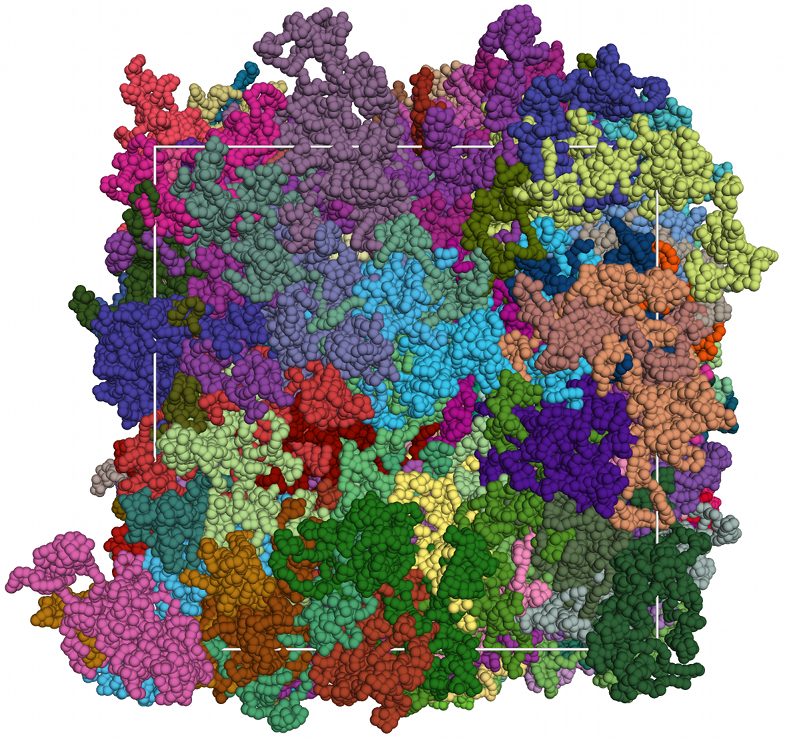}\\
  \includegraphics[width=0.7\columnwidth]{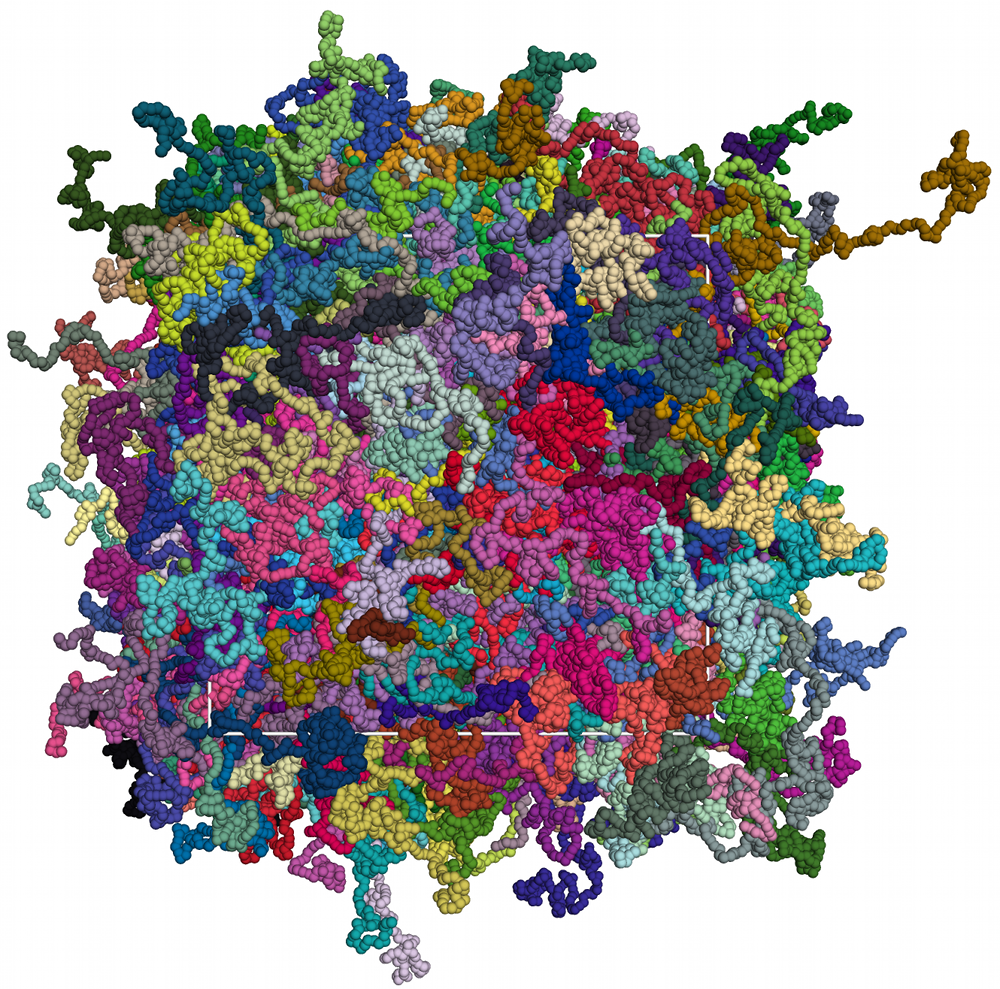}\\
  \caption{Upper image: Experimental image of chromosome
territories in a fibroblast cell nucleus. Image reproduced with
author’s permission from \cite{Cremer_Territories_Image}. This is a false color representation
of all of the chromosome territories visible in the mid-section of the
nucleus. Middle image: Territorial segregation in the model system
of nonconcatenated unknotted rings. The system in this image
consists of 200 rings of 1600 monomers each. Lower image: To
sharpen the statement that territorial segregation results from
topological constraints, we show here a system of 400 linear chains
of 800 monomers each. As the image demonstrates, the linear
chains mix completely, without any hint of segregation or territories.
This mixing is achieved via reptation and it is behind the Flory
theorem in polymer physics \cite{pgdg}. The middle and lower images
resulted from molecular dynamics simulation of an off-lattice model
(see \cite{Melt_of_Rings_Statics_2011} for details of the simulation and model}\label{fig:political_maps}
\end{figure}

As we already mentioned, mutual segregation or the incomplete penetration of rings provides a simple generic natural explanation of chromosome territories
\cite{Chromosome_territories_Cremer,Chromosome_territories_Cremer_Review_2006,Territories_Meaburn_Misteli_Nature_News_Views_2007}.
As a matter of fact, territorial segregation of chromosomes was seen also in HiC data \cite{HiC_Science_2009}.  But the most obvious view of territories was described in Refs.
\cite{Chromosome_territories_Cremer,Chromosome_territories_Cremer_Review_2006,Territories_Meaburn_Misteli_Nature_News_Views_2007}.
Recall that the microscopic image of an interphase nucleus, where each chromosome is stained a distinct color, looks like a political geographic map \cite{Chromosome_territories_Cremer} or contiguous set of colorful patches with irregular shapes, covering space (see also reviews
\cite{Chromosome_territories_Cremer_Review_2006,Territories_Meaburn_Misteli_Nature_News_Views_2007}).
In this sense, the melt of rings exhibits a very similar behavior. If each ring is colored in its own distinct color then a something of a map emerges as published on the back cover of ``Physics Today''\cite{PhysicsToday_Image}. This is illustrated in Fig. \ref{fig:political_maps}. We want to emphasize that no segregation, no territories are observed in the system of linear polymers. This fact is known in polymer physics as the Flory theorem \cite{pgdg}, and it is also illustrated by the lower panel in the Fig. \ref{fig:political_maps}. Thus, we see that topological constraints in a concentrated system of long polymers lead to territorial segregation.

This leads to the hypothesis that chromosomes do not intermix, do
not tangle, and remain distributed over their respective territories
mostly because of their topological constraints
\cite{Rosa_Everaers_PLOS_2008,Vettorel_Grosberg_Kremer_2009}. This
idea was also suggested in the works
\cite{Arsuaga_chromosomes,Stasiak_chromosomes},
where it was based directly on the topological repulsion
of nonconcatenated loops in a dilute system, without any attention to
the compact packing issue.
The role of
topological constraints was also emphasized in the experimental work
\cite{Marko_Topo_Constraints_2010}.

The initial formation of territory-like regions is also observed in
computer simulations and to a certain extent by scattering
experiments of polyelectrolyte gels and
solutions~\cite{Dobrynin1996,Micka1999,Spiteri2007,MannBA2011}. Upon
change in solvent quality or counterion valency the systems start
to shrink by chain contraction. This is facilitated by counterion
condensation along the backbone of the chains leading to complexes,
which still contain a (decreasing) net charge. At the early stages
of this process the so-called pearl necklace structure is observed.
Consequently these dense regions still repel each other, i.e., they
stay segregated. For perfect gels without any dangling chain ends
this segregation state seems to survive up to melt densities, while
solution studies of (short) linear polymers suggest a
relaxation into the Gaussian chain conformation once the density is
sufficiently high that the counterions are no longer localized on the
backbone and can diffuse freely throughout the melt or solution.

It is worth emphasizing the important differences in approach between
Refs.
\cite{Vettorel_Grosberg_Kremer_2009,Melt_of_Rings_Statics_2011},
\cite{Rosa_Everaers_PLOS_2008} and the simulation part of
Ref. \cite{HiC_Science_2009}. In
\cite{Vettorel_Grosberg_Kremer_2009,Melt_of_Rings_Statics_2011}, a
system of rings was considered to enforce topological constraints.
This system was carefully equilibrated. That is, the authors made
sure that the entropically dominant or statistically most probable
set of conformations was observed and studied. By contrast, the authors
of \cite{Rosa_Everaers_PLOS_2008} simulated (mostly) long linear
chains for a pre-reptation time, therefore, not even attempting to
equilibrate their system. What they have done instead is to take
advantage of having mapped the parameters of the simulated model
onto those of a real chromatin fiber. Accordingly, they were able to
argue that their molecular dynamics run was over a biologically
reasonable amount of time over which the interphase
stage of the cell cycle exists in
nature. These authors showed also that their results for linear
chains on a pre-reptation time were similar to those of the rings.  It
is interesting to realize that some sort of territorial segregation
was observed in three rather different simulation schemes.
In \cite{Rosa_Everaers_PLOS_2008} the authors start with compact unknotted
crumpled-like polymers and run the simulation much shorter than the
reptation time. In the simulation part of \cite{HiC_Science_2009}
the authors start from a very open coil conformation, force it to
collapse very fast, and then run it for a short time. While in
\cite{Vettorel_Grosberg_Kremer_2009,Melt_of_Rings_Statics_2011}
dense systems of rings were equilibrated, which is to say it could
be run indefinitely long. The similarity in the results
of these three very different
systems is a powerful demonstration of the validity of the very idea
that topological constraints are the key to segregation.

This idea is also supported by the study undertaken in a different
context by Vettorel and one of us (KK)
\cite{Vettorel_Developing_Entanglements}.  In that work, the authors
simulated melts of long collapsed polymers and found that relaxation
into the entangled state is significantly delayed to times
\textit{larger} than the reptation time once the chain lengths reach
of the order of about $50 L_e$. As shown above, typical chromatin
fibers exceed this length significantly.

We should repeat here the arguments why we think topological constraints to be so important for the chromatin fiber. Of course, the chromatin fiber is not a ring. Nevertheless, un-crossability is likely to play a huge role.  The simplest argument is that the cell life time is by far not sufficient to realize complete relaxation via reptation \cite{Rosa_Everaers_PLOS_2008,Sikorav_Jannink_1994}. Also, telomere regions at the ends of chromosomes are likely to form bulges which are not conducive for
reptation. The additional factor which slows down the reptation is that the chromatin fiber is probably attached in some places to the nuclear envelope (its inner surface). If that is the case then regular reptation becomes problematic and the situation becomes reminiscent of star-branched macromolecules, where the relaxation time is controlled by the arm
retraction mechanism and is exponentially long \cite{RubinsteinColby}.

We should also comment here about the role of topological enzymes
(topoisomerase) which are capable of cutting both strands of dsDNA,
passing another piece of DNA through the cut, and then gluing the cut back
together. This seems to make dsDNA effectively phantom or
self-crossable \cite{Sikorav_Jannink_1994}. Simple estimates show,
however, that the amount of enzymes in the nucleus and ATP (free
energy) supply in the cell is by far insufficient to forget about
topological enzymes. Nevertheless, this subject remains somewhat
controversial in the literature, as emphasized recently, e.g., by
H. Schiessel \cite{Schiessel_2013}. Sure enough, topoisomerase
enzymes play their role, but mostly in the metaphase. In fact,
apparently there is no indication of topoisomerase activity during
the interphase \cite{Mirny_private_communication}. In any case, the
activity of topological enzymes does not cancel the fact that to the
first approximation the chromatin fiber is not crossable
\cite{Rosa_Everaers_PLOS_2008}. Thus, it is sensible to hypothesize
that instead of using the sophisticated machinery of topological
enzymes to make many random crossings, nature would make good use
of polymer un-crossability to control territorial segregation.

Territorial segregation of chromatin fibers appears to be a common
feature of higher eukaryotes, including humans. Lower eukaryotes,
such as yeast seem to have much less pronounced territories, or no
territories at all
\cite{Territories_Meaburn_Misteli_Nature_News_Views_2007,HiC_Yeast,Zimmer_Group_Yeast}.
Given the relatively short genome of yeast (see Table
\ref{tab:examples_of_organisms}), this observation is at least
qualitatively consistent with our above estimates of the number of
entanglement lengths of the yeast genome.
It is interesting to mention that yeast has about as many genes (and
about as large genes) as humans. In other words, the amount of coding
DNA is nearly the same for both yeast and human. However, yeast
has almost no introns while we have plenty. This is why
human DNA is two orders of magnitude longer than that of yeast.

Since the qualitative picture of political maps agrees rather well for the model system of a melt of rings and for real chromosomes in cell nuclei, it makes sense to look at more details for both cases.

\section{``Territorial polymers'' and chromosome territories: quantitative
considerations}\label{sec:territories_quantitative}

\subsection{Modern experimental techniques of chromatin
investigation}\label{sec_sub:Modern_experimental_techniques}

In recent years, quantitative experimental information about
chromatin organization in space was delivered mostly by two
groups of methods, called ``FISH'' and ``3C'' with derivatives up to
HiC.  Recall that FISH basically measures $r(s)$, the
subchain size, while HiC measures $P(s)$, the loop factor or contact
probability.  We will discuss these results in greater detail below
in this section, but it is useful to remind briefly the basic ideas
behind these two methods.

The essence of the FISH method \cite{FISH_BOOK} is to label a few loci
on the genome with small fluorescent molecules. The spatial distance
of the fluorescing spots is then measured under a microscope. Recent advances in microscopy increase the resolution significantly beyond the wavelength of the light used (see, e.g., \cite{Cremer_FISH_2011,Cremer_FISH_2012}). By
doing this for many pairs of loci, one can -- ideally -- relate the
genetic distance between the loci $s$ or, in polymer physics parlance, the contour distance between
the monomers to their spatial Euclidean
distance $r(s)$ \cite{Neer_1977, FISH_Gerhard_Hummer_1,
FISH_Gerhard_Hummer_2, FISH_Yokota, FISH, FISH_Gilbert,
FISH_Jhunjhunwala, Cremer_FISH_2011, Cremer_FISH_2012}.

The central idea of the ``C'' methods is to chemically cross-link at
some particular time pieces of the chromatin fiber which
happen to be close to each other in space at that very moment. Then the
DNA is chopped into pieces by a restriction enzyme, and (omitting
important technical details here!) the parts which happened to be
cross-linked are sequenced. Cross-linking is usually realized by
formaldehyde which connects histones, but not DNA. Thus the
cross-linking acts on the level of the $10 \ \mathrm{nm}$ fiber, not
on the level of DNA. The cross-linked parts are then separated, for
instance, in the HiC version by using biotin labeled ends, followed
by deep sequencing. In any case, since the entire genome sequence is
known, and since the experiment is done on many cells in parallel,
the result allows one to reconstruct the contact map or, in polymer
physics language, the probability $W$ of contact between any two
monomers of the same chain $s^{\prime}$ and $s^{\prime\prime}$, say
$W(s^{\prime},s^{\prime\prime})$. It is usually assumed that on
average $W$ depends only on the distance $s = s^{\prime} -
s^{\prime\prime}$, and so one can find the so-called loop factor,
equivalently known as contact probability, $P(s)$ which is the probability
that two monomers a contour distance $s$ apart meet in space.

There is also one work \cite{Chromosome_Fractality_Scattering} which carried out
small angle neutron scattering experiments on the entire cell nucleus
(of a chicken erythrocyte). The global spatial
arrangement of the chromatin strands were analyzed. We comment
on this experiment later in Section \ref{sec:scattering_experiments}.

Since the essential quantitative information at our disposal comes
from measuring subchain sizes $r(s)$ and contact probabilities
$P(s)$, we start with a discussion of a simple approximate relation
between these two quantities.

\subsection{Mean field relation between subchain size $r(s)$ and loop factor $P(s)$}\label{sec_sub:1/s}

For any fractal conformation with the subchain size scaling as $r(s) \sim b s^{\nu}$, the loop factor can be estimated by the following argument.  Let us imagine that we hold one end of the subchain fixed in space. The second end is located at a distance of about $r(s)$, i.e., it is dispersed over the volume of the order of $r^3(s)$. Therefore, the probability, $P(s)$, to find it in a small volume $v \sim b^3$ around the first end is estimated as
\be P(s) \simeq v/r^3(s) \ . \label{eq:Mean_Field_relation_between_P_and_r} \ee
This assumes that the second end is more or less uniformly distributed over the volume $r^3(s)$, i.e, we make a mean field argument. Given that $r(s) \sim s^{\nu}$, this argument yields for the critical exponent $\gamma$:
\begin{equation} \gamma = 3 \nu \ . \label{eq:relation_of_gamma_nu} \end{equation}
For the Gaussian coil (with $\nu=1/2$) this gives the well known correct answer $\gamma=3/2$ in 3D (or $d/2$ in dimension $d$).  For the crumpled fractal globule $\nu=1/3$ (\ref{eq:crumpled_prediction}) and, therefore, $\gamma=1$.
This simple theoretical estimate is in excellent agreement with HiC
data for both human \cite{HiC_Science_2009} and mouse chromosomes
\cite{HiC_for_mouse}.

However nice the agreement between theory and experiment is, the
exact $1/s$ scaling for the loop factor cannot be correct, at least
not over an infinite range of length scales $s$. Take the total
number of neighbors of any given monomer, namely the sum $ \sum_s
P(s) $, which diverges for $P(s) \sim 1/s$, i.e., a monomer would
have a diverging number of close neighbors. For a chain of length
$N$, this would mean that the number of spatial neighbors for a given
monomer becomes of order $\ln N$. This is surely impossible in the
limit $N \to \infty$.

Nevertheless, from a theoretical point of view it is important to understand whether we are talking about a real fractal, which remains self-similar over an infinite range of scales and which approximates real chromatin over some finite range, or whether there is even theoretically no real fractal. Aesthetically the first alternative is more attractive, of course. But even logically, we can imagine arbitrarily long rings in the melt, much longer than any DNA in any real genome, and we want to understand the loop factor for
such system and find out whether this system is fractal or not. Thus, to summarize, the mean field estimate $P(s) \sim 1/s$ (or $\gamma = 1$), which is based on the assumption of statistical independence between the subchain ends within the volume $r^3(s)$, cannot be accurate. There should be some correlation between subchain ends for any $s$.

\begin{figure}[h] \centering
\includegraphics[width=0.45\textwidth]{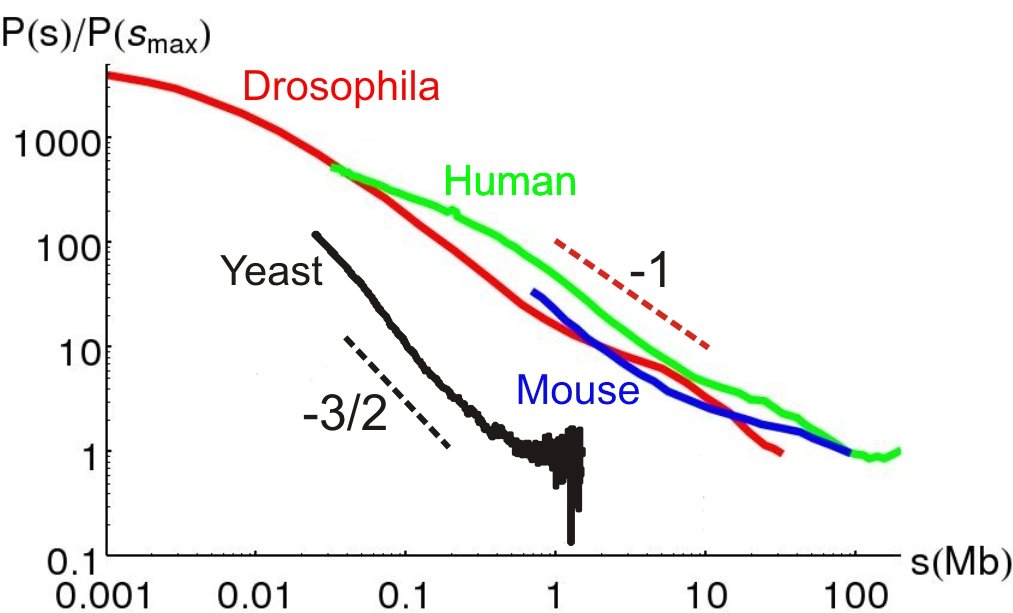}
\caption{HiC data of contact probability $P (s)$ for various organisms plotted in log–log scale against the genomic distance s. Every data set is normalized by the last data point. That means each curve  represents $P (s)/P (s_{max})$, where of course the maximal $s$-value, $s_{max}$, is different for different curves. In other words, every curve is vertically shifted such that its right-most point is on the level $1$. Solid red: Drosophila Chromosome 3R \cite{HiC_drosophila_Sexton}; solid blue: mouse \cite{HiC_for_mouse}; solid green: human \cite{HiC_Science_2009}; solid black: yeast \cite{HiC_Yeast}, analysis of the data in \cite{Zimmer_Group_Yeast_Predictive_Model}. Data for yeast, mouse, and human are averages over all chromosomes. Dashed red: slope $−1$ (corresponds to the mean field prediction $P(s)\sim s^{−1}$); dashed black: slope $−3/2$ (corresponds to equilibrium globule)}
\label{fig:HiC_all_animals}
\end{figure}

\subsection{Contact probability (loop factor) of chromatin:
experiments}\label{sec_sub:contact_probability}

The contact probability of chromatin $P(s)$ is the probability that
two DNA loci on the same chromosome, separated by the genetic
(contour) distance $s$ (measured, for instance, in the number of
base pairs), will be neighbors in real 3D space.  This probability
and its dependence on $s$ was measured for human
fibroblasts \cite{HiC_Science_2009},
mice \cite{HiC_for_mouse},
fruit flies \cite{HiC_drosophila_Sexton}, and yeast \cite{HiC_Yeast}.
All these data are collected in Fig.
\ref{fig:HiC_all_animals}.  We see that the contact probability
behaves rather similarly for human, mouse, and drosophila cells.  In
particular, \cite{HiC_Science_2009} reports that the loop factor
scales as $P(s) \sim s^{-\gamma}$ in the interval of roughly $0.5 \
\mathrm{Mbp} \lesssim s \lesssim 7 \ \mathrm{Mbp}$, with a
``critical exponent'' $\gamma = 1.08$ for the human genome. In a
similar interval, the data for mouse \cite{HiC_for_mouse} indicate
$\gamma = 1.05$. However, the implicit error bar of these
measurements is hard to estimate. This scaling appears to be
consistent with the crumpled globule model, based on the previous mean
field estimate of Eq. (\ref{eq:relation_of_gamma_nu}), as also emphasized
in \cite{Rosa_Everaers_Looping2010}. Based on this the authors of
the experimental work \cite{HiC_Science_2009} claimed agreement with
the theoretical predictions based on crumpled globules.

We leave aside the questions of error bars associated with the above
mentioned values of $\gamma$ as well as the question of the range of $s$
where the power law fitting is considered. We refer the interested
reader to the original papers as well as review articles
\cite{Mirny_Chromosome_Res_Review,Mirny_CurrentOpinion_2012,Dekker_Chapter_7}.
Instead, we concentrate on the following observations. First, yeast
definitely belongs to a separate class compared to all other
organisms presented in Fig. \ref{fig:HiC_all_animals}.  Second,
the contact probability for yeast is roughly consistent with equilibrium
globule behavior, both because the slope is close to $-3/2$ and
because of the leveling-off at larger $s$, at least for the largest
chromosome. Third, the contact probability for human, mice, and
drosophila is definitely inconsistent with the equilibrium globule, both
in terms of the slope being nowhere near $-3/2$ and the definite lack of
leveling off.

The situation with yeast was significantly clarified in the recent
work \cite{Zimmer_Group_Yeast_Predictive_Model}, where the authors
computationally examined an elaborate, but straightforward polymer
model of the yeast nucleus. The estimated values for the Kuhn segment and linear
density of yeast
chromatin fiber were $l_K = \unit[60]{nm}$ and
$1/b = \unit[83]{bp/nm}$. They took into account the real
lengths of all chromosome arms, the fact that each chromosome has a
centromere attachment, and the fact that chromosome ends (telomere
regions) are preferably located next to the nuclear envelope.
Importantly, they simulated the \textit{equilibrium} properties of the
system, and obtained a very good account of the contact probability
as well as a number of other structural data. The agreement
thus obtained gives a convincing proof that folding of chromatin
fiber in yeast can be reasonably considered as an equilibrium state
of a confined polymer. In particular, this explains why the contact
probability in yeast, as seen in Fig. \ref{fig:HiC_all_animals},
behaves very much like that of an equilibrium globule.

Fig. \ref{fig:HiC_all_animals} demonstrates beyond a doubt that
the chromatin of other species is organized differently compared to
yeast. Our main hypothesis, already stated several times in
different ways, suggests that the main difference is due to topological
constraints. The relatively short genome of yeast does not need to
reptate and, therefore, is equilibrated, as confirmed by the above
discussed conclusions of the work
\cite{Zimmer_Group_Yeast_Predictive_Model}.  We argue that genomes
of higher eukaryotes do not have enough time to reptate.  This of
course goes along with our assertion that chromosome territories
have topological origin.  We now continue and hypothesize that
the contact probabilities of the higher eukaryotes seen in Fig.
\ref{fig:HiC_all_animals} do not follow the equilibrium globule
behavior also for the same topological reason.  This
topology-controlled deviation from standard equilibrium globule
behavior is a manifestation of non-equilibrium of very long linear
chains, or in a ring model it may be viewed as an equilibrium
manifestation of topology.

\subsection{Subchain sizes in chromatin}\label{sec_sub:subchain_sizes}

\begin{figure}[h] \centering
\includegraphics[width=\columnwidth]{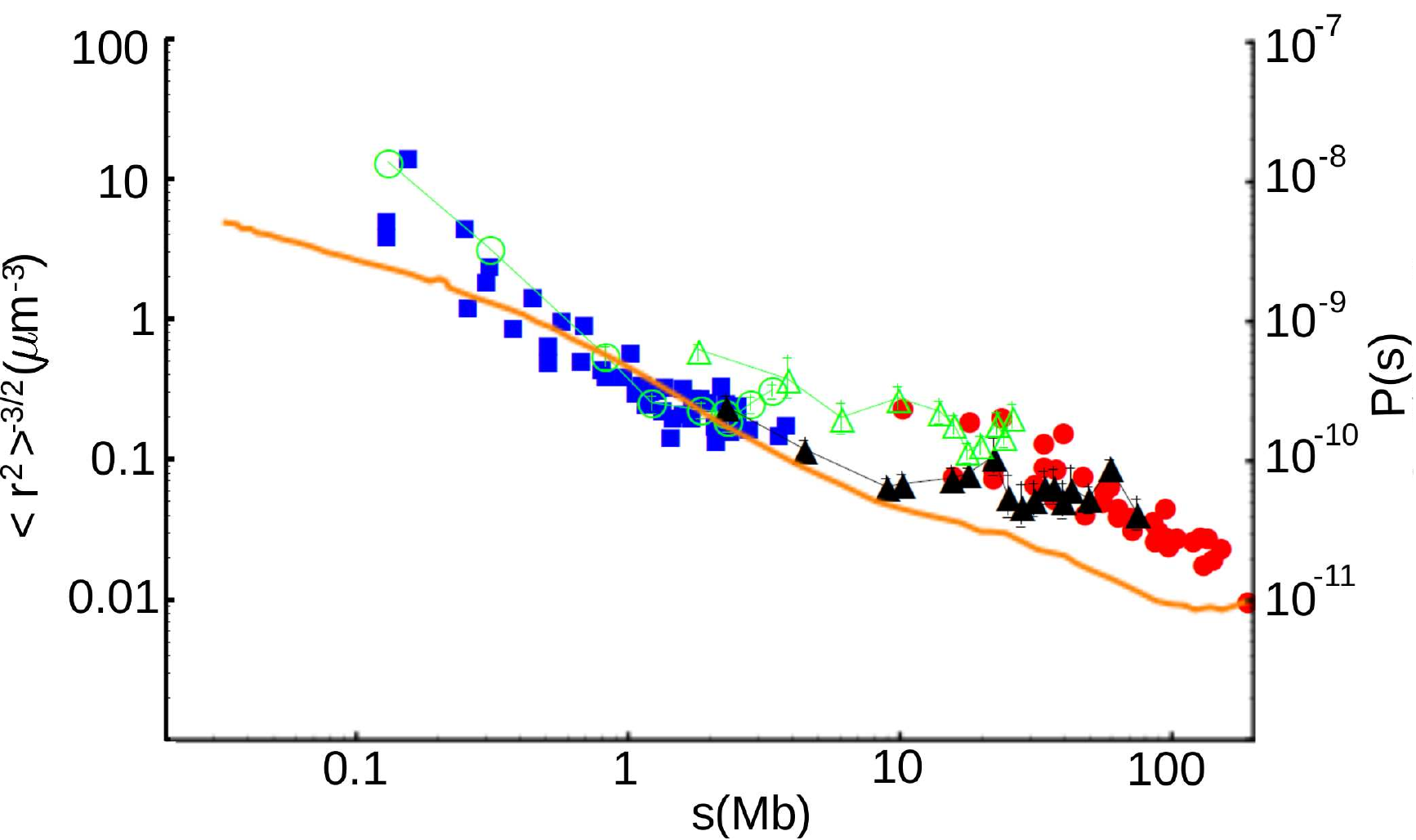}
\caption{
Collation of FISH (subchain size $r(s)$) and HiC (contact
probability $P(s)$) data, plotted log–log against the subchain length
$s$. In light of the mean field relation $P(s)\sim r^{−3}(s)$ (\ref{eq:Mean_Field_relation_between_P_and_r}), FISH data
are presented as $r^{−3}(s)$. All data are for human cells. The thick
orange line represents HiC data from \cite{HiC_Science_2009}. All other data are FISH:
blue filled squares \cite{van_den_Engh04091992}; red filled circles \cite{FISH_Yokota}; black filled
triangles \cite{Mateos-Langerak10032009} for chromosome 11; green empty triangles \cite{Mateos-Langerak10032009} for
chromosome 1; green empty circles \cite{Mateos-Langerak10032009} for ridges of chromosome
1. Where error bars are not shown, authors of the experiments claim
the error bars are smaller than the symbols used here. Note that of
the different estimates of the blob size, the value derived from
scattering experiments \cite{Chromosome_Fractality_Scattering} is about the largest. These authors quote
about \unit[300]{nm}, which corresponds to an inverse volume of about
$\unit[74]{\mu m^{−3}}$, indicating a maximal distance well below all distances
discussed here.}
\label{fig:FISH_HiC_human}
\end{figure}

The FISH method, which measures the subchain sizes $r(s)$, was
developed significantly earlier than the ``C'' methods. From the polymer physics
prospective, the value of $r(s)$ is also well known for regular
equilibrium globules: it behaves as $r(s) \sim s^{1/2}$ up to $s
\sim N^{2/3}$, i.e., up to the distance where the chain can cross the
available volume by a random walk. Then $r(s)$ levels off at
$r(s) \sim N^{1/3}$ and remains independent of $s$ at all larger
$s$.

The first FISH data were consistent with the Gaussian scaling $r(s)
\sim s^{1/2}$. For values of $s$ up to about $5 \times 10^{5}$ base
pairs the FISH data for $r^2(s)$ were satisfactorily approximated as
a straight line in coordinates $(r^2,s)$.  This tempted the authors
of Ref. \cite{van_den_Engh04091992} to formulate the ``random walk''
model of chromatin organization.  In fitting the data there was a
significant element of uncertainty for any analysis beyond the bare
power law. Namely, if we believe that chromatin is a Gaussian
polymer on a certain length scale, with a Kuhn segment of length
$l$, if the number of base pairs per unit contour length is
$\rho$ then $r(s) \simeq (l s/\rho)^{1/2} $ because $s/\rho$ is
the contour length. Although the Kuhn segment of bare DNA
is known to be about $100 \ \mathrm{nm}$, it is not easy to
determine \textit{a priori} the relevant Kuhn segment for chromatin
fiber, so the practical approach is to view it as an adjustable
parameter. FISH data, apart from the very scaling $r(s) \sim
s^{1/2}$, allow only to determine the combination of parameters
$l/\rho$, not each of them separately. Nevertheless, more
detailed FISH data for larger $s$ showed some sign of slowing down
in the growth of $r(s)$ which was interpreted as the signature of
spatial confinement. Accordingly, the work \cite{Hahnfeldt15081993}
introduced the model of a random walk in a confined volume. From a
polymer physics point of view, this means that Refs.
\cite{van_den_Engh04091992,Hahnfeldt15081993} interpreted the FISH
data as pointing to an equilibrium globule organization of
chromatin. The difficulties of making this model consistent
with excluded volume constraints, topological constraints, and
the lack of reptation relaxation were not addressed in the
early papers \cite{van_den_Engh04091992,Hahnfeldt15081993}.  Also, a
fully developed saturation of $r(s)$ was not observed. Therefore, an
alternative interpretation of the slowing down of the growth of
$r(s)$ might be appropriate.

A more recent FISH study \cite{Mateos-Langerak10032009} makes the
point that $r(s)$, at least for some chromosomes,
saturates at a value of $s$ significantly smaller than the nucleus
size, as is clearly visible in Fig.
\ref{fig:FISH_HiC_human}.  This suggests some mechanism of
saturation other than volume confinement. The authors of
\cite{Mateos-Langerak10032009} argue that this is evidence of
what they call loop formation.
They refer then to the well known fact of entropic (topological
in nature) repulsion between closed polymeric loops in dilute
solution, as first reported in the work
\cite{Maxim_second}.  We note in passing that the territorial
segregation of rings in a concentrated system, which is the central
subject of our work, can be viewed as the result of the same
repulsion in a very dense system.  Importantly, this repulsion
exists only for loops which maintain their closeness and remain
nonconcatenated over the entire longevity of the experiment, i.e., in
physics language, the loops have to be quenched.  Annealed loops, which
close and open in the course of thermal motion, exhibit no repulsion
at all.  Meanwhile, the fact that two monomers $i$ and $j$ gave a
signal in a particular Hi-C experiment suggests that the piece of
the chain between them formed a ``loop'' at the time of the experiment,
but it does not mean this loop was quenched before the experiment,
\textit{in vivo}.  In our opinion, the interpretation of each Hi-C
signal as an indication of a real quenched loop in the system is a
mistake.

Another interpretation would relate this effect to the territorial segregation.  Namely,  we can
expect that a smaller chromosome gets topologically confined among bigger ones, just as one ring
in the melt gets compressed by its neighbors.  Such an explanation would of course make sense if
and only if the size of even a small chromosome significantly exceeds the entanglement length
$N_e$.  That is one of the reasons why we discussed the estimate of $N_e$ in detail before.
According to the summary given in Table \ref{tab:entanglement_length_estimates}, even the sizes
of small chromosomes, whose $r(s)$ saturates at distances $s$ of
about $\unit[0.5]{Mbp} = \unit[500]{kbp}$ (cf. Fig. \ref{fig:FISH_HiC_human}), are long enough for the topological mechanism to be
functional. The overall dependence of $r(s)$ being at least roughly consistent with the expected
crumpled globule scaling $r(s) \sim s^{1/3}$ (\ref{eq:crumpled_prediction}) is visible in
Fig. \ref{fig:FISH_HiC_human} in the $s>N_e$ region.

In recent years, much more detailed models have been developed to fit the
FISH data.  In most cases, the authors talk about delicately organized
and balanced loops supported and controlled by special long lived
cross-links (see, for instance,
\cite{Ostashevsky01111998,Ostashevsky01062002,Mateos-Langerak10032009,Arya_PRE}).
The idea is that these cross-links are created by special
biological mechanisms.  There is no way for us to negate such a
hypothesis. At the same time, one should first exhaust all
possibilities to explain observations based on simple robust physics
before resorting to biological machinery and the special actions of
such machinery.
Indeed, it seems that nature is
``opportunistic'' in the sense that it employs simple physics
whenever possible. Although this question can perhaps be
argued in different ways, here we concentrate on the possibility to
explain observations regarding the chromatin organization based on
topological constraints and polymer physics.

\subsection{Theoretical issues: contacts between subchains}\label{sec_sub:theoretical_issues}

Let us return for a moment to the mean field relation $\gamma = 3 \nu$
(\ref{eq:relation_of_gamma_nu}), and to the discussion of the fact that this relation cannot be exact for a true fractal over the unrestricted range of scales.  The latter conclusion means that the critical exponent $\gamma$ is not determined by the value of $\nu$, i.e., it is a different, independent exponent.
The existence of critical indexes independent of $\nu$ is not so surprising given our experience in polymer physics \cite{desCloizeaux_Jannink_book,pgdg,RedBook,RubinsteinColby}.

Thus, for the rings in the melt we have to turn our attention to exponents independent of $\nu$.  That means, we should take a more detailed view of conformations. One useful perspective opens up if we look at surfaces between neighboring rings in the melt of rings, or surfaces between subchains, i.e.,   to consider the index $\beta$ which describes the roughness of the boundaries of territorial polymers. Furthermore, there are two closely related surface exponents $\beta_1$ and $\beta_2$. They are defined in the following way:

\begin{itemize} \item Consider an $s$-long part of the chain.  How many contacts does
this part have with all other monomers?  This scales with index $\beta_1$:
\be \# \mathrm{all \ contacts \ except \ with \ itself} \sim s^{\beta_1} \ .
\label{eq:beta_1_definition} \ee
This index characterizes the ``surface area'' of the blob.
\item Consider two blobs of length $s$
each, located next to one another in space. How many contacts do they have? This scales with
index $\beta_2$:
\be \# \mathrm{contacts \ with \ another \ blob} \sim s^{\beta_2} \ .
\label{eq:beta_2_definition} \ee
This index describes the ``contact area'' between two touching blobs.
\end{itemize}

The relation between indexes $\beta_1$ and $\beta_2$ has to do with the number of blobs $Q$ which
can simultaneously be neighbors to one another:
\be s^{\beta_1} = Q s^{\beta_2} \ . \ee
To estimate $Q$ we can resort to the physics of polymer solutions (see, e.g., book \cite{pgdg}).  Each subchain of length $s$ is spread over volume $\sim s^{\nu d}$ (in $d$ dimensions). This volume can contain $s^{\nu d}$ monomers, but since each subchain has $s$ monomers we arrive at
\be Q \sim \frac{s^{\nu d}}{s} = s^{\nu d -1} \ \label{eq:overlap_scaling} \ee
which yields
\be \beta_1 = \beta_2 + \nu d -1 \ . \label{eq:relation_between_beta_1_and_beta_2} \ee
For the dense fractal system with $\nu = 1/d$ we get $\beta_1=\beta_2$, which is why we used simply $\beta$ in the work \cite{Melt_of_Rings_Statics_2011}, deriving the relation
\be \gamma + \beta = 2 \ . \label{eq:relation_between_beta_and_gamma} \ee
Given that there are two indices $\beta_1$ and $\beta_2$, there are two relations:
\bea \gamma + \beta_1 & = & \nu d +1 \ \label{eq:relation_between_gamma_and_beta_1} \\ \gamma +
\beta_2 & = & 2 \ . \label{eq:relation_between_gamma_and_beta_2}
 \eea
Because of Eq. (\ref{eq:relation_between_beta_1_and_beta_2}) these two relations are
equivalent, and each of them gets reduced to the familiar result
(\ref{eq:relation_between_beta_and_gamma}) in the case $\nu=1/d$.  Below we briefly outline
arguments leading to the relations (\ref{eq:relation_between_gamma_and_beta_1}) and
(\ref{eq:relation_between_gamma_and_beta_2}).  The arguments in Sections
\ref{sec_subsub:counting_argument} and \ref{sec_subsub:blob_contact_argument} were developed
during a discussion with R.Everaers and M.Rubinstein
\cite{Everaers_Grosberg_Rubinstein_KITP_discussion}.

\subsubsection{Counting argument \cite{Everaers_Grosberg_Rubinstein_KITP_discussion}}\label{sec_subsub:counting_argument}

Consider a polymer chain which is a space-filling fractal (i.e., $\nu = 1/d$), and suppose its contact probability is $P(s)$, with large $s$ asymptotics being $s^{-\gamma}$, $\gamma>1$.  For one monomer, the number of ``foreign'' contacts (i.e., with monomers a distance $s$ or greater away along the chain) is proportional to $\sum_{s^{\prime}=s}^{\infty} P(s^{\prime})$, and for the
whole subchain this number is $s$ times larger.  Thus
\bea \# & \mathrm{all} & \mathrm{contacts \ except \ with \ itself}  \sim  s^{\beta_1} \nonumber \\
& \sim & s \times \sum _{s^{\prime}=s}^{\infty} \left. s^{\prime} \right.^{-\gamma} \sim
s^{2-\gamma} \ , \eea
which is the result (\ref{eq:relation_between_gamma_and_beta_1}).

\subsubsection{Self-similarity argument}

Here we paraphrase the original argument given in \cite{Melt_of_Rings_Statics_2011}. Let us count how many pairs of monomers, a distance $s$ apart along the chain, are in contact.  There are $\sim N$ pairs a distance $s$ apart each, and with the probability of contact for each pair being $s^{-\gamma}$, the number of pairs is $Ns^{-\gamma}$.

Let us now re-count this same number in a different manner by using blobs of some $g$ monomers each. We have $\frac{N}{g}$ pairs of blobs, and the probability of contact between blobs is $\left(\frac{s}{g} \right)^{-\gamma}$, yielding $\frac{N}{g} \times \left(\frac{g}{s} \right)^{\gamma}$ blob contacts.  Each blob contact delivers $g^{\beta_2}$ monomer contacts. However, these monomer contacts are between monomers a chemical distance $s \pm g$ apart, and only the
$1/g$ fraction of those are a distance $s$ apart. Thus, the number of monomer contacts is now expressed as $\frac{N}{g} \times \left(\frac{g}{s} \right)^{\gamma} \times g^{\beta_2} \times \frac{1}{g} $.  This must coincide with the original estimate $N s^{-\gamma}$ because the number of monomer contacts cannot depend on how we count them. By equating the two
\be \frac{N}{g} \times \left(\frac{g}{s} \right)^{\gamma} \times g^{\beta_2} \times \frac{1}{g} = \frac{N}{s^{\gamma}} \ee
one arrives at formula (\ref{eq:relation_between_gamma_and_beta_2}).

\subsubsection{Blob contact argument \cite{Everaers_Grosberg_Rubinstein_KITP_discussion}}\label{sec_subsub:blob_contact_argument}

The probability of contact between two monomers a distance $s$ apart, for instance, $t-s/2$ and $t+s/2$, is $s^{-\gamma}$. Consider now two subchains of length $s$ each, namely, one from $t-s$ to $t$ and another from $t$ to $t+s$.  They are centered at the two monomers of interest.  These two subchains have $s^{\beta_2}$ contacts.  What is the probability that one of these contacts is between the monomers of interest, $t-s/2$ and $t+s/2$?  There are $\sim s^2$ possible choices of pairs of contacting monomers, therefore, the probability that one of these pairs will be chosen among the $s^{\beta_2}$ actual pairs is proportional to $s^{\beta_2}/s^2$. Thus this argument yields
\be \frac{s^{\beta_2}}{s^2} = \frac{1}{s^{\gamma}} \ , \ee
which is again the result (\ref{eq:relation_between_gamma_and_beta_2}).

\subsubsection{Examples and comparisons}

It is instructive to see how all the relations above work in various known cases.  First, consider a melt of linear chains.  In this case, $\nu = 1/2$, $\gamma = d/2$, $\beta_2 = \frac{4-d}{2}$ (controls contacts between two overlapping blobs), and $\beta_1 = 1$ (controls contacts of a given chain with monomers of any other), so all relations above work perfectly.

It is worth emphasizing that the relations
(\ref{eq:relation_between_gamma_and_beta_1}) and
(\ref{eq:relation_between_gamma_and_beta_2})
represent general properties of self-similarly organized polymers. They do not rely on any specific properties of crumpled globules. In particular, they are not based on the mean field estimate of $\gamma =\nu d$. Since we know already that the mean field estimate is not accurate, we need to make our considerations not assuming anything specific about the value of $\gamma$.

For the ``naive'' crumped globule, best exemplified by a Hilbert curve or its circular analog called Moore curve \cite{Space_Filling_Curves_Book} in 3D (Fig. \ref{fig:Hilbert_curve}), we have $\nu = 1/d = 1/3$ and also all surfaces are pretty smooth, so $\beta_1 = \beta_2 = 2/3$ and then $\gamma = 4/3$.

By contrast, if we accept the mean field estimate $\gamma= \nu d$, Eq.
(\ref{eq:relation_of_gamma_nu}), and then take $\nu = 1/d$ for a dense fractal globule, we get $\beta_1=\beta_2 = 1$ and also $\gamma=1$. That means in this approximation, the surfaces of a fractal globule include some finite fraction of all monomers. In fact, these surfaces are so ``rough'' that they can hardly be called surfaces. Quite apart from the terminological question of whether we can call
different rings with $\beta = 1$ to be territorially segregated, we point out that such structures cannot be fractal with an unlimited spectrum of scales, which is the same conclusion we made based on the logarithmic divergence of the number of contacts.

In general, there are several obvious conditions on the indexes: in $d$ dimensions $1/d \leq \nu \leq 1$ and $1-1/d \leq \beta_1 \leq 1$, while $\beta_2 \leq \beta_1$, and $0 \leq \beta_2 \leq 1$.  It follows then from the relations above that in general $1 \leq \gamma \leq 2$, while specifically for the compact case $\nu=1/d$ a stricter bound holds: $1 \leq \gamma \leq 1+1/d$.

The numerical results for the melt of rings \cite{Comparing_Lattice_and_off-Lattice} indicate quite complex finite size behavior of different quantities.  The total number of surface monomers for
either the full ring or its subchains demonstrates a very accurate power law behavior throughout the studied range of polymer length, and shows $\beta_1=0.97$. This value is just below unity, so one is tempted to suspect that it is unity.  However, the fraction of surface monomers does decrease with chain length, so it is indeed possible that $\beta_1 < 1$.  Even more
convincing data suggesting $\beta_1<1$ are obtained from the scaling of the static structure factor, as we discuss below.  At the same time, the contacts between blobs exhibit a very different finite size behavior, for the seeming power $\beta_2$, i.e., the slope of a $log-log$ plot of monomer-monomer contacts between blobs against blob length $s$ increases monotonically from well
below $0.5$ to somewhere close to unity at the largest $s$ accessed in simulation so far.  At the same time, the contact probability $P(s)$ exhibits complementary behavior with the seeming value of $\gamma$ (in the same sense) increasing with $s$, while the number of blob neighbors $Q(s)$ seems
to saturate at large $s$ in accordance with (\ref{eq:overlap_scaling}). The nature of this finite size behavior remains to be understood.

\subsection{Static structure factor and scattering off
chromatin}\label{sec_sub:static_structure_factor}

\subsubsection{Spatial nuclear density (in-)homogeneity: scattering
results}\label{sec:scattering_experiments}

The whole line of quantitative arguments above relies on the assumption that there are no strong
density variations on scales of the order of the entanglement length and beyond. The classical
color pictures of chromatin territories (cf. Fig. \ref{fig:political_maps}) suggest a homogeneous
density throughout the cell nucleus. This however might be misleading as the wavelength of
visible light is too large to easily allow one to resolve greater local density variation by eye. To be
more accurate, one could resort to scattering experiments. For that we can start from the
structure of the ``semidilute chromatin solution'', which can in principle be determined by
scattering experiments \cite{pgdg,RubinsteinColby}.

In a melt or dense solution there are no density fluctuations expected to exist beyond the scale
of the blob size. On scales smaller than the blob diameter, one can resolve the single chain
structure. This, however, requires sufficient scattering contrast between the chromatin fiber and
its immediate surrounding, i.e., typically salty water and proteins. However, even within the
scale of the blobs, this scattering contrast usually is very weak. For X-rays to be used
efficiently the electron density is too small and for neutrons the overall density of H atoms
does not display enough variation. To improve this, one can create an artificial contrast. For
instance, in the case of neutron scattering of polymer systems, the usual procedure is deuteration,
i.e., replacement of hydrogen atoms for deuterium in one of the components of the system. Here
this would correspond to a (partial) deuteration of the water. Neutron scattering is a perfect
tool to study subtle variations in H and D densities, since H and D have dramatically different
scattering lengths \cite{Borsali2008,Richter2000}.

We are aware of only one piece of experimental work \cite{Chromosome_Fractality_Scattering}
examining chromatin structure (of a chicken erythrocyte) by small angle neutron scattering (see
also the work \cite{Chromosome_Fractality_Scattering2} by the same group attempting to simulate the data
). In this experiment, scattering indeed was based on the
contrast between DNA and surrounding water, a mixture of $\mathrm{H_2O}$ and $\mathrm{D_2O}$. On
length scales within the blob size one expects a variation of the scattering intensity as a
function of momentum transfer, providing information about the structure of the chromatin fiber.
For larger scales both the density of the solvent and chromatin are averaged out, suppressing
any change in the scattering signal. From this crossover one can directly read off the blob
size, at least in principle. The scattering experiment \cite{Chromosome_Fractality_Scattering}
indicates a homogeneous smeared out DNA density on length scales above $300$ to $\unit[450]{
nm}$ or, equivalently, for momentum transfers between $q\leq 2 \pi/\unit[450]{nm}$ and $q\leq 2
\pi/\unit[300]{nm}$. While this confirms the concept of a homogeneous, concentrated solution, the
quoted blob size certainly is too large to accommodate the contour length of the chromatin fiber
with $l_p \approx \unit[150]{nm}$ and about $\unit[40]{bp/nm}$. Indeed, standard
polymer physics arguments \cite{pgdg,RedBook,RubinsteinColby} imply that the blob size is about $\xi =
\left(\rho_K l_K \right)^{-1/2}$ (because $\xi < l_K$ implies that the fiber is nearly straight inside the
blob). For the parameters of the human cell nucleus discussed above, this would be about $\xi \approx
\unit[60]{nm}$. For the higher density chicken erythrocyte $\xi$ should be even smaller.  This
discrepancy between measured blob size and theoretical estimates can result from different
sources. One possible reason is due to the overall small accuracy of the data because the
system is much more complex than a regular semidilute polymer solution, where such methods have
been employed very successfully. In addition, experiments have been performed on the nuclei of chicken
erythrocytes, which are quite different from other eukaryotic cells.

The experiments of Ref. \cite{Chromosome_Fractality_Scattering} also
provide the first information about the fractal property (on scales
below the blob diameter) of the DNA fold itself, but cannot reveal
any information of the global indices $\nu$, $\beta$, or $\gamma$.
It also does not give information about the relevant contour length
and the Kuhn segment of the chromatin fiber inside the blob, which
would provide an independent estimate of the entanglement length.

\subsubsection{Scaling of the static structure factor}

Although a few scattering experiments have been performed so far, this is a very promising technique,
which motivates us to discuss what we can expect theoretically.  In general, the static structure
factor is the most informative characteristic of conformations. We denote it $S(q)$, with $q$
the scattering wave vector ($q= 2\pi \sin \vartheta/2/\lambda$, where $\vartheta$ is the scattering angle
and $\lambda$ is the wavelength of scattered waves). Usually, in the most interesting interval
of $q$, namely $2\pi/R_g < q < 2\pi/b$, the static structure factor scales as $S(q) \sim
q^{-1/\nu}$. Accordingly, one would expect to be able to distinguish a Gaussian coil ($S(q) \sim
q^{-2}$) or a ball with smooth surface (Porod law, $S(q) \sim q^{-4}$) from a crumpled globule
with expected scaling $S(q) \sim q^{-3}$.

This conclusion, however appealing, holds only on scales below the
blob size, and its application is tricky because of the fractal
surfaces between territories, which smear out the scattering
contrast. Corresponding analysis was presented in
\cite{Melt_of_Rings_Statics_2011} based on the ideas of the works
\cite{Duplantier,Strasbourg,CarmesinKremer_1990} for 2D melts. The
argument was originally given without making the distinction between
$\beta_1$ and $\beta_2$ by assuming that they were the same. The essence of
the argument can be formulated in the following way.

Imagine that we labeled (for instance, deuterated) some $n$ monomers. The static structure factor is defined by the formula:
\begin{equation} S(q) = \frac{1}{n} \sum_{i \neq j}^{n} \exp \left( \imath \mathbf{q} \cdot
\left( \mathbf{r}_i - \mathbf{r}_j \right) \right) \ . \label{eq:structure_factor_definition}
\end{equation}
For the moment let us assume that the labeled part is either one entire ring ($n=N$) or some part of it ($n<N$). In this case, the static structure factor (\ref{eq:structure_factor_definition}) has the following properties. First, at $q=0$ we always obtain $S(q=0)=n$. Second, in the intermediate range of $q$
we expect some power law dependence $S(q) \sim q^x$, where the power $x$ is to be found. Moreover, since the overall size $R \sim b n^{\nu}$ is the only relevant length scale (since we deal with one ring or its part), we can write $S(q) \sim n (q R)^x \sim q^x n^{1+\nu x}$. Third, and this is the most delicate part of the argument, the only place where scattering can take place is the
surface of the labeled part. Therefore, the total scattered intensity, which is equal to $n S(q)$, must depend on the number of labeled monomers as $n^{\beta}$. Comparing, we conclude that $2 + \nu x = \beta$. Therefore, the structure factor scales as
\begin{equation} S(q) \sim \frac{n^{\beta - 1}}{q^{(2 - \beta)/\nu} } \ .
\label{eq:structure_factor_scaling} \end{equation}

For instance, for the linear chain in a regular polymer melt all monomers belong to ``surface'',
as they contact with other chains, so $\beta=1$ and we return to the familiar result $S(q) \sim
q^{-1/\nu}$ (where $\nu=1/2$ for the melt).  But for the melt of rings $\beta \neq 1$ and the use
of a more sophisticated scaling (\ref{eq:structure_factor_scaling}) is necessary. Computational data for both lattice and off-lattice systems were fit using formula (\ref{eq:structure_factor_scaling}), and the result $\beta = 0.93$ was found \cite{Comparing_Lattice_and_off-Lattice}.  The fitting procedure used in our works \cite{Melt_of_Rings_Statics_2011,Comparing_Lattice_and_off-Lattice} was criticized and placed under doubt in a recent comment \cite{Strasbourg_comment} (see also our reply in \cite{Our_reply_to_Strasbourg_comment}). However, it is worth emphasizing that the scaling formula (\ref{eq:structure_factor_scaling}) was not questioned, only the numerical fitting procedure.  In light of this criticism, extracting reliable estimates of $\beta$ and $\gamma$ from static structure factor measurements will have to wait for more accurate data.  Nevertheless, our present understanding, which is based on the several different measurements discussed above, remains that $\beta$ is slightly smaller than unity 
while $\gamma$ is slightly larger than one.

\section{Dynamics}
\label{sec:Dynamics}

From a comprehensive polymer physics viewpoint, we should consider the dynamics  of the ring melts and not only the statistics of equilibrium conformations.  This is a difficult task, because reptation \cite{pgdg,DoiEdwards,RubinsteinColby,RedBook,GiantMolecules}, the workhorse of polymer dynamics theory, does not apply for the rings as they have no ends.

The simplest model, the dynamics of a single ring in a lattice of
immobile obstacles, can be analyzed in great detail (see, e.g.,
the problems for Chapter 9 in the book
\cite{RubinsteinColby} as well as papers
\cite{Rubinstein_PRL_1986,Obukhov_Rubinstein_Duke_PRL_1994}).
However, since the lattice of obstacles concept is not proven even
for statics, its usefulness for dynamics remains questionable.
Recent works \cite{Rubinstein_talk_in_Leiden,
Generalized_Lattice_Animal, Generalized_Lattice_Animal_Dynamics} may
change this situation and give more credence to the generalized
lattice animal model based on annealed branched polymer theory.  It
is perhaps too early to decide how successful this theory is.

So far we have to rely on simulations. We can summarize the main findings
\cite{Melt_of_Rings_Dynamics_2011} as follows:

If $\mathbf{r}(t)$ is the time dependent position vector of one ring (say, its mass center, with
the convention that $\mathbf{r}(t=0)=0$), then at very long times $\left< \mathbf{r}^2(t) \right>
\simeq 6 D_N t$, and we can address the $N$-dependence of the diffusion coefficient. It was found
\cite{Melt_of_Rings_Dynamics_2011} that $D_N \sim N^{-2.3}$, which is not dramatically different
from the result of similar simulations and experiments for linear chains, $D_N \sim N^{-2.4}$. At
the same time the viscosity of the melt of rings increases with $N$ dramatically slower than
for linear chains, as $\eta \sim N^{1.4}$ (instead of $N^{3.4}$ for linear chains).

At smaller times, before the cross-over to regular diffusion, $\left< \mathbf{r}^2(t) \right>$
was found to exhibit sub-diffusion, although somewhat faster sub-diffusion (roughly $t^{3/4}$ for
the ring sizes considered in the simulations) than for linear chains ($t^{1/2}$).

For linear chains the cross-over between the sub-diffusion and diffusion regimes for the displacement of the center of mass of the chains is observed at about
the Rouse time $O(N^2)$, well below the time of stress relaxation, $O(N^{3.4})$. Surprisingly, this is not the case for the ring systems, where molecules continue to sub-diffuse long after the stress has completely relaxed, as far as the accuracy of the present simulations can determine this. An explanation of this observation, beyond the fact that shape fluctuations do not necessarily contribute to the overall ring diffusion, remains to be found.

In terms of mapping polymer models to chromatin, an interesting estimate can be performed based on the fact that the diffusion coefficients behave nearly identically while viscosities behave so dramatically different for the melt of territorial globules versus a melt of entangled linear chains. This difference indicates that the relaxation mechanism, which controls viscosity, in the territorial system is decoupled from translational motion, which is manifested in the diffusion coefficient.  Therefore, the ratio of viscosities for territorial globules and for entangled linear chains, which scales as $\sim N^{-2}$ according to our computational result, reflects the ratio of the relevant relaxation times.  This ratio is about $2 \times 10^2$ for the longest chains for which both  linear and ring polymers were tested in \cite{Melt_of_Rings_Dynamics_2011} ($N=800$, $N/N_e\approx 29$). Therefore, simple extrapolation suggests that even for short human chromosomes, for which $N/N_e \approx 1200$, the relaxation 
time is shortened by about 5 orders of magnitude (a factor of about $3.2 \times 10^5$) because of territorial segregation. Obviously, this is a very significant advantage.

Much more attention to dynamics was paid by Rosa and Everaers in their work \cite{Rosa_Everaers_PLOS_2008}.

Another interesting aspect in which the melt of rings model can be compared to chromatin data has to do with sub-diffusion at shorter time scales.  We here only point out that most reported observations of sub-diffusion in the cell nucleus deal with rather small length scales, below about 200 nm \cite{Chromosome_Fractal_Review_Bancaud,Chromosome_fractality_experiment_Bancaud,Dynamics_Golding,Dynamics_Banks,Dynamics_Bronstein}, and it may be unrelated to the topic of our interest, which is the role of topological constraints. Or in polymer language, these phenomena are likely to happen below the concentration blob and tube diameter scales.

\section{Role of sequence}\label{sec:Sequence}

The fractal crumpled globule hypothesis is
attractive because it is based on very generic properties of
polymers and does not involve any assumptions of an evolutionary
developed special machinery. But we know of course that DNA does
have a sequence of base pairs, which cannot be ignored. It is a
particularly relevant and intriguing question in regards to the
non-coding DNA, the introns.  Their role in general is unclear, and
one cannot exclude the possibility that they are somehow involved in
the control of genome folding.  If there is any relation between large
scale features of sequence and structure, its casual nature can be
speculated to be in any of two directions. On the one hand,
since chromatin conformations are to some extent fractal for the
reasons of topology discussed above, it is possible to imagine DNA
sequences which in the course of evolution develop some degree of
self similarity. On the other hand, it is also possible that a
particular conformation or group of conformations of chromatin can
be additionally stabilized by selecting the proper DNA sequence,
provided that protein-mediated interactions are sufficiently complex
and diverse.

It is instructive to compare this situation with the relation between sequences and structures
known for proteins \cite{ShakhnovichReview}.  The use and applicability of these types of ideas for chromatin is in no way self-evident.  We can speculate that the role of volume interactions in
the chromatin case will be relegated to the complex relations between chromatin pieces realized via
histone and non-histone proteins and nucleosomes.  We point out in this regard that the
relation between fractal properties of sequences and fractal arrangement of conformations is known even for models with very simple interactions \cite{ShakhnovichReview,Govorun,Nechaev_HiC_Maps}; it seems, therefore, very reasonable to expect that much more sophisticated interactions, involving architectural proteins and the like, are also likely to lead to some sort of connections between sequences and structures.

\section{Concluding remarks}\label{sec:Conclusion}

The genome of a eukaryotic cell is stored in the cell nucleus as a very long chromatin fiber.  It
is confined rather densely in the limited space of the nucleus, but at the same time all (or at
least many) parts of it are accessible for cell machinery manipulations.  This means, chromatin
fiber is incomparably less tangled than a random thread of comparable length packed into a
comparably small volume. The underlying principles governing this efficient genome organization
are poorly understood.

We argue that at least a significant fraction of this puzzle belongs
to the realm of polymer physics.  This follows from a very generic
argument: chromatin fiber has a one dimensional connectivity and it
obeys the excluded volume constraint.  Therefore, its conformation
in space belongs to the class of so-called self-avoiding walks
(SAW). However, there are a huge variety of SAWs,
and this raises questions about the SAWs that chromatin follows and why they are found in the cell.

Two types of SAWs are most frequently considered in polymer physics, neither of which is a good candidate to model chromatin: isolated SAWs (dilute set of coils in a good solvent) or a dense tangle of SAWS (as in a melt). In both cases the SAWs are self-similar fractals, but different ones, characterized by different swelling exponents $\nu$ for their gyration radii ($R_g \sim N^{\nu}$ with $\nu \approx 3/5$ for an isolated SAW and $\nu = 1/2$ for the melt).  The third commonly considered type, the so-called equilibrium globule, is typically a single chain SAW confined in a volume comparable to the joint excluded volume of its monomers. This is also not a good candidate to model chromatin because it is very heavily tangled.

The selection of a specific type of SAW in real chromatin is controlled in two ways.  First, chromatin fiber is not naked DNA but a complex of DNA with a large variety of proteins \cite{Dekker_Chapter_7}. The pieces of such a complex ``polymer'' do obey the excluded volume, but they also interact in a complex way via their associated proteins. Second, the chromatin fiber is thermodynamically not an equilibrium system and its folding is under some sort of a dynamic control. Following the work \cite{crumpled2}, we hypothesize that these mechanisms select SAW conformations which are weakly tangled, because it is necessary for biological functions, but otherwise random.  In other words, the hypothesis is that conformations, although not equilibrated, correspond to conditional maximum of entropy, conditioned on weak tangling.

This leads then to the workhorse model of nonconcatenated rings, first suggested in
\cite{Rosa_Everaers_PLOS_2008}.  It cannot be overemphasized that we do not assume chromatin
fibers to be rings. The model of rings is motivated only by the above hypothesis of partial,
incomplete entropy maximization, restricted by weak tangling.  Of course, this is consistent with
the fact that the estimated relaxation time (``reptation time'') for the chromatin fiber is much
longer than the life time of any cell.

Because of this logic, we reviewed in this article two interconnected but distinct lines of research. On the one hand, there is the pure polymer physics problem: a melt of nonconcatenated unknotted rings.  It turned out to be a challenging problem in its own right.  On the other hand, we attempted to gain insight into chromatin based on the newly acquired knowledge about rings.
Our theoretical thinking in this direction was informed by the new generation of experiments, including those on chromosome territories, FISH, and ``C'' (from 3C to Hi-C)
\cite{van_den_Engh04091992,FISH_Yokota,Chromosome_territories_Cremer,3C,FISH_Gilbert,
4C,FISH_Jhunjhunwala,HiC_Science_2009,Marko_Topo_Constraints_2010,Cremer_FISH_2011,
HiC_for_mouse,HiC_drosophila_Sexton,Cremer_FISH_2012} as well as computational work
\cite{Rosa_Everaers_PLOS_2008,Everaers_2010_Gene_Colocalization,Arya_PRE,Zimmer_Group_Yeast_Predictive_Model,
Mateos-Langerak10032009}.

The results presented indicate that rings in the melt are segregated for topological reasons,
i.e., they correspond to SAWs with $\nu = 1/3$.  This gives a very natural explanation of the
phenomenon of chromosome territories.  Furthermore, conformations of rings in the melt appear to
be self-similar fractals consistent with the crumpled globule idea \cite{crumpled1}. The fractal
properties of chromatin were noted and discussed in a number of contexts
\cite{Chromosome_Fractal_Review_Bancaud, Garini_2012}. The melt of rings model sheds some light
on these observations.  More detailed experiments, such as Hi-C, address properties which from a
theoretical viewpoint go beyond the index $\nu$ and have to do with an independent set of
indices, denoted above as $\beta$ and $\gamma$.  Our theoretical understanding of the
corresponding properties is incomplete even on the level of rings.  Simulations indicate that
$\gamma$ appears to be slightly above unity, meaning that the contact probability between DNA
loci decays as $s^{-\gamma}$, somewhat faster than $s^{-1}$ which seems to be consistent with
observations for higher eukaryotes such as humans and mice \cite{HiC_Science_2009, HiC_for_mouse,
HiC_drosophila_Sexton}.  At the same time, Hi-C data for yeast are closer to $\gamma=3/2$ which
is characteristic of an equilibrated polymer globule and consistent with the fact that yeast chromatin is
significantly shorter, owing to the lack of non-coding DNA.

These successes seem sufficiently encouraging to continue the study of the topological properties of weakly tangled dense polymer systems with the idea in mind to learn more about chromatin.  This is why we would like to conclude by listing a set of open questions pertaining to both the melt of rings model
system and its more general application to chromatin. We start with
relatively simple, well-defined problems and then continue to increasingly tough questions.

It is established that one can construct space-filling curves with any desirable value of
$\gamma>1$. However, what are the entropically dominant values of the critical exponents $\beta$
and $\gamma$ for a polymer system without knots? In other words, what are $\beta$ and $\gamma$
for the majority of all non-tangled space-filling curves?  (We emphasize here that the powers $\beta$
and $\gamma$ are not very accurately measured in experiments and characterize a relatively narrow
window of scales; however, we view them as an important tool to understand the system, which is
why we view them as relevant.) How can one understand the sub-diffusion of rings in dense,
nonconcatenated systems and how can one understand the fact that sub-diffusion continues for a
time longer than the apparent stress relaxation time? How could this relate to the properties of
chromatin fibers in territories?  How does it relate to the observations of sub-diffusion (of
telomeres and of inserted particles) in chromatin?  What is the role of DNA sequence in the
formation and support of chromatin folding (via its effect on histones and other proteins in and
around chromatin fiber and thus on the fiber-fiber interactions)? In particular, what is the
connection between the fractal organization of chromatin in space and long-range correlations in
non-coding DNA sequences \cite{Stanley_correlations}?  (Although arguments in this direction in
the work \cite{crumpled2} were based on the naive assumption $\beta = 2/3$, the question remains
valid and open.) How do crumpled globules or other polymer models react to chemical ``signal''
modifications, such as methylation, which formally means perturbing the parameters of volume
interactions? How do particles diffuse through the chromatin matrix?

Another group of questions pertains to the bridge between polymer physics and chromatin folding. As we discussed, the physical parameters of chromatin fiber are poorly known at present, including the simple ones such as persistence length or even linear density (the number of base pairs per unit contour length).  We also pointed out that the entanglement length depends on these parameters, but also very sensitively on the density, which in this case is controlled by the cell nucleus size. The volume of the nucleus can easily be different by a factor of two for different cells. For nonentangled territorial chromatin fibers this factor only weakly affects the friction, and thus the prefactor to the relevant relaxation times.  For an entangled system the resulting change of $L_e$ would alter the relaxation times dramatically, asking for different relaxation mechanisms for different cells.  We view this as a powerful argument supporting our main
topological hypothesis.

This list of open questions is easy to continue. The important conclusion that we want to emphasize is that modern quantitative experimental methods allow one to place the serious physical analysis of chromatin folding on rather solid footing. The first impressive steps have been taken in the context of FISH and HiC experiments, to name the two most prominent approaches. As physicists, we expect scattering experiments to pick-up and to provide valuable information as well.

We also see grounds to expect a large role for simulations.  They appear to be extremely promising for a thorough analysis of the static and dynamic properties of well-defined model systems. They allow a direct comparison to either experiment or theory or both. In the case of melts of nonconcatenated rings, the simulations so far provide the only way to generate data of perfectly controlled, equilibrated systems. They impressively demonstrate the crucial role of topological constraints. Already a simple visual inspection illustrates the similarity of ring polymer territories and chromosome territories. The quite close agreement of numerical estimates of the exponent $\gamma$ from simulation data of long, concentrated ring polymers and those from ``C''  and ``FISH'' experiments, further supports our view that polymer physics concepts can provide very helpful arguments on our way to better understand the complex morphology of the chromatin fiber in the interphase cell nucleus.

\section*{Acknowledgements}

The authors acknowledge useful discussions with Robijn Bruinsma, Job
Dekker, Ralf Everaers, Yuval Garini, Gary Grest, Maxim Imakaev, Erez
Lieberman-Aiden, Ron Milo, Leonid Mirny, Sergei Nechaev, Rob
Phillips, Yitzhak Rabin, Michael Rubinstein, Jean-Louis Sikorav,
Jean-Marc Victor, Edouard Yeramian, Alexandra Zidovska, Christophe
Zimmer. This research was supported in part by the National Science
Foundation under Grant No. NSF PHY11-25915.  KK, JS, and AYG
acknowledge the hospitality of KITP Santa Barbara where part of this
work was completed.  KK acknowledges the hospitality of the Center
for Soft Matter Research at NYU and AYG acknowledges the support
from the Alexander von Humboldt Foundation through a Senior Scientist
Award to spend time at the MPI for Polymer Research. Research
carried out in part at the Center for Functional Nanomaterials,
Brookhaven National Laboratory, which is supported by the US
Department of Energy, Office of Basic Energy Sciences, under
contract no. DE-AC02-98CH10886.

\bibliography{Chromosome_References}

\end{document}